\documentclass[12pt]{iopart}
\usepackage{iopams}  
\expandafter\let\csname equation*\endcsname=\relax
\expandafter\let\csname endequation*\endcsname=\relaxD
\usepackage{epsf,graphicx,graphics}
\usepackage{amsmath}
\usepackage{latexsym,amssymb}
\usepackage{setspace,cite}
\usepackage{hhline}
\usepackage{booktabs}
\usepackage{a4}
\usepackage{slashed}
\usepackage{float}
\usepackage[normal]{caption}
\usepackage[left=3cm,right=3cm,top=3cm,bottom=4cm]{geometry}

\newtheorem{thr}{Theorem}

\newtheorem{lem}[thr]{Lemma}
\newtheorem{prop}[thr]{Proposition}
\newtheorem{cor}[thr]{Corollary}

\newcommand{\rar}{\rightarrow}

	\newcommand{\slim}{\sum\limits}
	
	\newcommand{\disfrac}[2]{\displaystyle{\frac{#1}{#2}}}
	
	\newcommand{\essu}{\mathfrak{su}(2)}
	\newcommand{\sun}{\mathfrak{su}(N)}

	\newcommand{\bv}[1]{\mbox{\textbf{#1}}}
	\newcommand{\bsym}[1]{\boldsymbol{#1}}
	
	\newcommand{\R}{\mathbb{R}}
	
	\newcommand{\C}{\mathbb{C}}
	\newcommand{\Z}{\mathbb{Z}}
	
	\newcommand{\mc}[1]{\mathcal{#1}}
	\newcommand{\mk}[1]{\mathfrak{#1}}

	\newcommand{\lnorm}{\langle\!\langle}
	\newcommand{\rnorm}{\rangle\!\rangle}
	\newcommand{\HIP}[2]{\langle\, #1\,|\,#2\,\rangle}
	\newcommand{\HHIP}[2]{\lnorm\, #1\,|\,#2\,\rnorm}
	\newcommand{\ccdot}{\!\cdot\!}
	\newcommand{\norm}[1]{\lVert#1\rVert}

	\newcommand{\labeq}[2]{\begin{equation}\label{#2} #1 \end{equation}}

\begin{document}

\title[Existence of spherically symmetric adS EYM theories with compact gauge groups]{On the global existence of spherically symmetric hairy black holes and solitons in anti-de Sitter Einstein-Yang-Mills theories with compact semisimple gauge groups}

\author{J. Erik Baxter}

\address{Dept. of Engineering and Maths,
	Sheffield Hallam University,
	Howard Street,
	Sheffield, 
	South Yorkshire S1 1WB}
\ead{e.baxter@shu.ac.uk}

\begin{abstract}
	We investigate the existence of black hole and soliton solutions to four dimensional, anti-de Sitter (adS), Einstein-Yang-Mills theories with general semisimple connected and simply connected gauge groups, concentrating on the so-called \textit{regular} case
	. We here generalise results for the asymptotically flat case, and compare our system with similar results from the well-researched adS $\mk{su}(N)$ system. We find the analysis differs from the asymptotically flat case in some important ways: the biggest difference is that for $\Lambda<0$, solutions are much less constrained as $r\rightarrow\infty$, making it possible to prove the existence of global solutions to the field equations in some neighbourhood of existing trivial solutions, and in the limit of $|\Lambda|\rar\infty$. In particular, we can identify non-trivial solutions where the gauge field functions have no zeroes, which in the $\sun$ case proved important to stability.
\end{abstract}

\pacs{04.20.Jb, 04.40.Nr, 04.70.Bw, 11.15.Kc}
\noindent{\it Keywords}: Hairy black holes, Solitons, Semisimple gauge group, Anti-de Sitter, Einstein-Yang-Mills theory, Existence \\
\section{Introduction}

Research into Einstein-Yang-Mills (EYM) theory, which concerns the coupling of gauge fields described by the Yang-Mills (YM) equations to gravitational fields described by Einstein's equations, has become abundant in the literature in the last couple of decades. This work began in considering asymptotically flat, spherically symmetric, `hairy' black holes \cite{bizon_colored_1990} and solitons (`particle-like solutions') \cite{bartnik_particle-like_1988}, coupled to a gauge field with structure group $SU(2)$. This field of enquiry first emerged in the 1980s and thus the asymptotically flat $\essu$ and $\sun$ systems are now well understood in a variety of cases -- see e.g. \cite{kleihaus_black_1998, vanzo_black_1997, kleihaus_su3_1995, kleihaus_charged_1997, kleihaus_su3_1995-1, mann_topological_1997}.

The problem with asymptotically flat EYM systems is that they have some tricky properties which provide analytical and numerical difficulties when obtaining solutions. First, global solutions are not abundant: due to strong constraints on the boundary conditions in the limit $r\rar\infty$, and at the origin in the case of solitons (see e.g. \cite{baxter_existence_2008}), regular solutions may only be found for certain discrete points in the boundary parameter space \cite{breitenlohner_static_1994, smoller_existence_1993, smoller_existence_1993-1, smoller_smooth_1991} and so global solutions are hard to find both numerically and analytically. Connected to this is their stability: $\sun$ purely magnetic solutions decouple into two sectors upon a linear perturbation, and spectral analysis shows that $\essu$ solutions possess $n$ unstable modes in each sector, where $n$ is the number of nodes (zeroes) of the gauge field; and in addition, these $\essu$ solutions must possess at least one node \cite{lavrelashvili_remark_1995, volkov_number_1995, mavromatos_aspects_1996, mavromatos_existence_1998}. This is related to the discrete nature of the globally regular solutions which are separated by continua of singular solutions: a small perturbation will turn any existing regular solution into a singular one. A node in the gauge field corresponds to a reversal of the field direction -- in a physical sense, we may intuit that this will lead to the instability of solutions. This instability result can be extended to general compact semisimple gauge groups, so that any global solutions that could be found would be necessarily unstable \cite{brodbeck_instability_1996}.

However for $\Lambda<0$, the picture changes completely. Here, because of the `box-like' geometry of anti-de Sitter (adS) space, it is much easier to set up the `balancing act' occurring between the repulsive YM forces and the attractive force of gravity, whereas for $\Lambda\geq 0$, the geometry is `open' and hair will in general destabilise and radiate away to infinity or else collapse inwards. It can be shown that in the adS case, we in general get a continuum of solutions in the parameter space \cite{breitenlohner_non-abelian_2004, bjoraker_stable_2000, bjoraker_monopoles_2000, winstanley_existence_1999}, making them much easier to find and to analyse. Connected to this, we may also find nodeless solutions, and can show that at least some of these are stable in the cases of $\essu$ for spherically symmetric \cite{winstanley_existence_1999} and non-spherically symmetric \cite{sarbach_linear_2001, winstanley_linear_2002} perturbations. Also we have established linear stability for $\sun$ spherically symmetric \cite{baxter_stability_2015} and so-called `topological' \cite{baxter_existence_2015} solutions. For a review of recent solutions, see \cite{winstanley_menagerie_2015}.

Furthermore, adS solutions have been considered recently for other applications: due to the adS/CFT (Conformal Field Theory) correspondence, gravitational theories in the bulk of adS space can be translated into particle theories on the boundary, meaning that results concerning hairy black holes (in particular) may provide insight into Condensed Matter Physics (CMP) phenomena (for a review of adS/CFT holography, see \cite{witten_anti_1998}).

Quite recently, the literature has been replete with special cases of hairy solutions in adS EYM theory, including cases such as \emph{dyons} (possessing a non-trivial electric sector of the gauge potential) \cite{nolan_existence_2012,nolan_stability_2016,baxter_existence_2016}, and topological black holes \cite{baxter_topological_2016} of the kind first considered in \cite{van_der_bij_new_2002}. This work has solely considered the gauge group $SU(N)$. However, in the case of asymptotically flat, spherically symmetric solutions with a \emph{general} compact gauge group and for the case of the so-called \emph{regular} action (defined in \cite{oliynyk_local_2002} and referred to as `generic' in \cite{brodbeck_generalized_1993} -- see Section \ref{sec:FEs}), it is found that the field equations are very similar to the $\sun$ case, and many qualitative features of the solutions carry over as well \cite{oliynyk_local_2002}.

Therefore, it seems logical to perform the same experiment on the asymptotically adS, spherically symmetric EYM system for a general compact semisimple gauge group, and to see how many features are present in both the general case and the specific $\sun$ case. Also strongly motivating this work is the possibility of exploring a very wide class of matter theories, both for the sake of CMP, and for further refinement of the ``no-hair'' theorem (see Section \ref{GGEConc}) which is relevant to gravitational physics. For the regular case at least, which is the main case considered in the literature so far, we see that it is not even necessary to know the YM one-form connection explicitly in order to obtain the field equations -- all the information one needs is essentially in the Cartan matrix of the Lie algebra of the structure group $G$ which represents the gauge field, making it easy to apply to a wide spectrum of EYM theories.

The outline of this paper is as follows. First, in Sections \ref{sec:conn} and \ref{sec:FEs} we will describe how we use our ans\"{a}tze to carve down the general field equations for four dimensional adS EYM theory with a general compact gauge group in the case of the `regular action', which we will describe later; and we show that in doing so, it coincides with the \emph{principal} action -- this allows us to simplify the field equations considerably. In fact, they become very similar in form to the field equations for $\sun$ \cite{baxter_existence_2008}. In Section \ref{sec:BCs}, we consider the boundary conditions needed for our solutions to be regular at $r=r_h$ (or $r=0$) and as $r\rar\infty$. In Section \ref{sec:asym}, we examine the asymptotic limit of the field equations $r\rar\infty$ in a `dynamical systems' sense, which turns out to be much simpler than it was for asymptotically flat space. Then in Section \ref{sec:embed} we identify some trivial embedded solutions, which are important to our final results.

In Section \ref{sec:locex}, we prove the existence of solutions locally at the boundaries, which are unique and analytic in their boundary parameters. Finally, in Section \ref{sec:gloex}, after proving that solutions may be regularly integrated out from the initial boundary into the asymptotic regime, we finish by establishing our main results: that global nodeless black hole and soliton solutions may be found in a neighbourhood of some trivial solutions found in Section \ref{sec:embed}, which are everywhere regular and uniquely and consistently specified by their boundary conditions; and that nodeless black hole and soliton solutions can be found in the limit $|\Lambda|\rar\infty$ (Section \ref{sec:l0}), anticipating a later investigation into the stability of these solutions. In Section \ref{GGEConc} we present our conclusions.

\section{Spherically symmetric, purely magnetic Yang-Mills connections for asymptotically adS spacetime}\label{sec:conn}

For asymptotically flat space, it is found \cite{oliynyk_local_2002} that we can reduce our attention from considering all possible conjugacy classes of bundle automorphisms by restricting focus to those for which the YM fields decay sufficiently fast at either boundary ($r\rar\infty$, and/or $r=0$ if the solution is a soliton). These are called `regular models' in \cite{bartnik_structure_1997} and correspond to the `zero magnetic charge' case in \cite{brodbeck_self-gravitating_1994}. A conjugacy class of $SU(2)$ bundle automorphisms is characterised by a generator $W_0$ which is an element of the Cartan subalgebra $\mk{h}$ -- for regular models, $W_0$ must be an $A_1$-\emph{vector}, i.e. the defining vector of a $\mk{sl}(2)$-subalgebra of $\mk{g}$. There is a remarkably wide variety of such actions for the case of $\sun$, as noted by Bartnik \cite{bartnik_structure_1997}; and such $A_1$-vectors are finite and have been tabulated \cite{dynkin_semisimple_1952,malcev_commutative_1945}.

The presence of a non-zero $\Lambda$ does not \emph{directly} affect the automorphism classes on the bundle structure, and therefore some similar results to \cite{oliynyk_local_2002} will here be derived, as we describe how to express the field equations for these regular models. But $\Lambda$ does make a difference asymptotically, and so we find a big difference in the regularity requirements for solutions in the limit $r\rar\infty$ (as may be expected from previous treatments of $\sun$ \cite{baxter_existence_2008}); as such, we note that the definition of `regular models' as given above must be amended a little for asymptotically adS space.

Let $G$ from here on be a compact semisimple connected and simply connected gauge group with Lie algebra $\mk{g}$. To consider spherically symmetric EYM connections is to consider principal $SU(2)$ automorphisms on principal $G$-bundles $E$ with base manifold $M$ (our spacetime), such that the automorphisms project onto isometry actions in $M$ whose orbits are diffeomorphic to 2-spheres. Since there is no natural action of $SU(2)$ on $E$, we must consider all conjugacy classes of such automorphisms. These conjugacy classes are in one-to-one correspondence to integral elements $W_0$ of a closed fundamental Weyl chamber $\overline{W(\Sigma)}$ belonging to a base $\Sigma$ of the roots of $\mk{g}$ with respect to a chosen Cartan subalgebra $\mk{h}$ \cite{bartnik_structure_1997, bartnik_spherically_1989, brodbeck_generalized_1993}.

Let $\mk{g}_0$ be the (real) Lie algebra of the structure group $G$ of the bundle $E$, so that $\mk{g}=(\mk{g}_0)_\C$, its complexification. Also, let $\{\tau_i\}$, $i\in\{1,2,3\}$ be the standard basis of $\essu$ defined using the Pauli matrices, with commutator relations $[\tau_i,\tau_j]=\epsilon_{ijk}\tau_k$, for $\epsilon_{ijk}$ the Levi-Civita antisymmetric symbol. Then $W_0$ may be chosen such that
\begin{equation}
W_0=2i\lambda(\tau_3),
\end{equation}
where $\lambda$ is the homomorphism from the isotropy group $\mc{I}_{x_0}$ of the $SU(2)$-action on $M$ at the point $x_0\in M$, determined by
\begin{equation}
k\ccdot\pi_0=\pi_0\ccdot\lambda(k),\quad\forall k\in\mc{I}_{x_0}\mbox{  if  }\pi_0\in\pi^{-1}(x_0),
\end{equation}
where $\pi^{-1}(x_0)$ is the fibre above $x_0$ and the central dot notation denotes the adjoint action.

The subject of possible classes of connections over principal bundles has been covered in the literature by Wang and others \cite{wang_invariant_1958, kobayashi_foundations_1963, kunzle_analysis_1994}. For instance, it is known that we may write the metric in common spherical Schwarzschild-type co-ordinates $(t,r,\theta,\phi)$ as 
\begin{equation}\label{metric}
ds^2=-\mu S^2dt^2+\mu^{-1}dr^2+r^2\left(d\theta^2+\sin^2\theta d\phi^2\right).
\end{equation}
%
%
%
Note that we here consider only static solutions, meaning all field variables are functions of $r$ alone.

In addition, Brodbeck and Straumann \cite {brodbeck_generalized_1993} show that in this case a gauge may always be chosen such that the Yang-Mills one-form potential is locally given as
\begin{equation}\label{conn}
\mc{A}\equiv A_\mu dx^\mu=\tilde{A}+W_1d\theta+\left(W_2\sin\theta+W_3\cos\theta\right)d\phi.
\end{equation}
In the above, $\tilde{A}$ is a one-form defined on the quotient space of the manifold which is entirely parametrised by the $(t,r)$ co-ordinates, representing the `electric' part of the connection. Here we consider the \textit{purely magnetic} case, and hence we set $\tilde{A}\equiv0$. We note that for $\Lambda=0$ this sector is not available in regular models \cite{oliynyk_local_2002}; it is available for $\Lambda<0$ but we find in the $\sun$ case that the condition $\tilde{A}=0$ still yields a rich space of solutions \cite{baxter_existence_2008}.

Also, we have $W_3=-\frac{i}{2}W_0$ as the constant isotropy generator, and we have constraints on $W_1$, $W_2$ (both also functions of $r$),
\begin{equation}\label{WE1}
[W_3,W_1]=W_2,\qquad [W_2,W_3]=W_1,
\end{equation}
which we refer to as the \textit{Wang equations} \cite{wang_invariant_1958}. 

However, we still have a countably infinite number of possible actions of $SU(2)$ on $E$: one for each element in $\overline{W(\Sigma)}\cap I$, the intersection of the closed fundamental Weyl chamber and the integral lattice defined by $I\equiv\ker(\exp|_\mk{h})$. Now for regular models, we require the YM fields to be non-singular at the centre $r=0$ (for solitons) and asymptotically as $r\rar\infty$.

In the case of $\Lambda=0$, this implied that
\begin{equation}\label{hom0}
[\Omega_1^0,\Omega_2^0]=W_3,
\end{equation}
and/or
\begin{equation}\label{hominf}
[\Omega_1^\infty,\Omega_2^\infty]=W_3,
\end{equation}
where we define 
\begin{equation}
\Omega_i^k\equiv\lim_{r\rar k}W_i(r)
\end{equation}
for $i\in\{1,2\}$, $k\in\{0,\infty\}$. That is to say, for asymptotically flat space, in at least one of these limits (if they exist) there has to exist a Lie algebra homomorphism from $\essu$ into $\mk{g}_0$; and if both limits exist, there also must exist a homomorphism between $\Omega^\infty_i$ and $\Omega^0_i$.

The reason for the constraints \eqref{hom0} and \eqref{hominf} is that in asymptotically flat space, the values of the gauge field functions $\omega_j$ at $r=0$ and as $r\rar\infty$ (taken in a particular basis that we will describe) must be equal to a particular set of constants $\{\lambda_j\}$ that depend on the Cartan matrix of the reduced subalgebra in question. This implies that the soliton solutions have no magnetic charge, according to \cite{brodbeck_self-gravitating_1994}. The constraints on the boundary values of the gauge fields are necessary so that the tangential pressure $p_\theta$ and energy density $e$ (see Section \ref{sec:FEs}) remain regular at infinity. 

However, for $\Lambda<0$ we have a different scenario. As we shall see, the values of the gauge field functions at the centre $r=0$ are still highly constrained, reflecting the singular nature of that boundary, and thus \eqref{hom0} still holds; but asymptotically, the ``fall-off'' conditions required to force the gauge field to be regular are much laxer than for $\Lambda=0$, and thus the gauge field functions and their derivatives will in general approach arbitrary asymptotic values. Again this is due to the nature of the asymptotic system considered in a dynamical systems sense. 

Our investigation in Section \ref{sec:asym} will show that this lack of asymptotic constraints on the YM field is to do with the nature of the variable change that we perform to render the asymptotic field equations autonomous, which in the case of asymptotically flat space necessitates the trajectory of every regular solution to end at a critical point (which we'll call $\Omega_i^*$, $i=1,2$) in the phase plane of the system. The critical points of the field equations are thus $\omega_j^{* 2}=\lambda_j$ for $j=1,...,\mc{L}$, where $\mc{L}=\mbox{rank}(\mk{g})$; the important point here being that for $\Lambda=0$, one is forced to have $\Omega_i^\infty\equiv\Omega_i^*$ ($i=1,2$), whereas for $\Lambda<0$, $\Omega_i^\infty\neq\Omega_i^*$ ($i=1,2$) in general. 

Hence, \eqref{hominf} does not have to hold for our solutions, and as we will see, this is manifested in the fact that for adS space, no constraints are placed on the gauge field functions or their derivatives as $r\rar\infty$, and we are allowed solutions with a global magnetic charge fixed essentially by the Cartan matrix of the reduced subalgebra, for which the tangential pressure and the energy density remain regular asymptotically. (Of course, \eqref{hom0} and \eqref{hominf} will both be trivially satisfied by embedded Schwarzschild anti-de Sitter solutions (see Section \ref{sec:embed}), and so for this solution at least, there must also exist a Lie algebra homomorphism from $\Omega_i^0$ into $\Omega_i^\infty$ \cite{oliynyk_local_2002}.) It must be noted though, it is still obviously true from the field equations that for regularity we must have

\begin{equation}\label{homcrit}
[\Omega_1^*,\Omega_2^*]=W_3.
\end{equation}

Thus, for asymptotically adS space, the \textit{system itself} still will possess the constraints \eqref{homcrit} at the critical point $\Omega^*_i$, but solutions will not reach the critical point of the system in general, freeing the asymptotic solution parameters from the constraints that are seen in the $\Lambda=0$ case. This is what is responsible for the much larger space of black hole solutions in the $\sun$ case, which we see need obey \emph{neither} \eqref{hom0} nor \eqref{hominf}; though we also emphasise that at the origin, regular solutions must still obey \eqref{hom0}. Thus, as in the case of $\sun$ for adS, we may expect the local existence proofs to be straightforward for $r=r_h$ and $r\rar\infty$ and much more involved at the origin $r=0$.

Now since $W_3$ is constant, \eqref{hom0} and \eqref{homcrit} represent constraints also on $W_3$, and hence on $W_0$ which must be the generating vector of an $A_1$- (i.e. $\mk{sl}(2)$-) subalgebra of $\mk{g}$. However the set of such so-called \textit{$A_1$-vectors} is finite, and have been tabulated by Dynkin \cite{dynkin_semisimple_1952} and Mal'cev \cite{malcev_commutative_1945} using what they call ``characteristics'', which are in one-to-one correspondence with finite ordered sets of integers chosen from the set $\{0,1,2\}$. These strings of integers then represent the value of the simple roots on $W_0$, the defining vector of the $A_1$-subalgebra, chosen so that it lies in $\overline{W(\Sigma)}$; and the tables of Mal'cev and Dynkin therefore give us a classification of all possible spherically symmetric, purely magnetic EYM models which obey the correct regularity conditions asymptotically and at the centre, for any compact semisimple simply connected gauge group.

\section{Field equations in the case of the `regular' action}\label{sec:FEs}

To proceed, we can note that out of all the possible actions classified by Dynkin and Mal'cev \cite{dynkin_semisimple_1952,malcev_commutative_1945}, these exists a privileged class of actions which corresponds to a principal $A_1$-vector in Dynkin's terminology, which Oliynyk and K\"{u}nzle \cite{oliynyk_local_2002} called \emph{principal} actions. There exists a slightly larger class of actions called `regular' in \cite{oliynyk_local_2002} (and `generic' in \cite{brodbeck_self-gravitating_1994}), for which the defining vector lies in the \emph{interior} of a fundamental Weyl chamber. (The other \emph{irregular} case involves the defining vector being on the boundary of a Weyl chamber.)

In this Section we will show that for $\Lambda<0$, as it was for $\Lambda=0$, all models with a regular action can be reduced to those with the principal action, for any semisimple gauge group. In terms of the field variables, this means that the YM potential can be chosen to be composed of real functions due to a gauge freedom, and that there are $\mc{L}$ of such functions where $\mc{L}=\mbox{rank}(\mk{g})$. We also have two metric functions governed by the Einstein equations: $m$ (the \emph{mass} function) and $S$ (the \emph{lapse} function). Then the field equations are determined by $\mc{L}+2$ real functions of the radial co-ordinate $r$ alone (for static, spherically symmetric solutions), and possess singularities at the centre $r=0$, the event horizon $r=r_h$ and as $r\rar\infty$.

A more convenient basis to use here for the Wang equations \eqref{WE1} in place of the generators $W_1$ and $W_2$ is
\begin{equation}\label{WD}
W_\pm=\mp W_1-iW_2,
\end{equation}
in which case equations \eqref{WE1} become
\begin{equation}\label{WE2}
[W_0,W_\pm]=\pm 2W_\pm, \qquad [W_+,W_-]=W_0.
\end{equation}
Then $W_\pm(r)$ are $\mk{g}$-valued functions, $W_0$ is a constant vector in a fundamental Weyl chamber of $\mk{h}$, and $\{W_0,W_\pm\}$ is a standard $\essu$ triple in the limit $r=0$ and at the critical points of the system. Also, $\mk{h}$ is the Cartan subalgebra of the complexified form of the Lie algebra, i.e. $\mk{h}=\mk{h}_0+i\mk{h}_0$, for $\mk{h}_0$ the real Cartan subalgebra of $\mk{g}_0$, which in turn is the real compactified form of $\mk{g}$. Naturally, we introduce a complex conjugation operator $c:\mk{g}\rar\mk{g}$ with convention
\begin{equation}
c(X+iY)=X-iY,\,\,\forall X,Y\in\mk{g}_0.
\end{equation}
This implies that
\begin{equation}
W_-=-c(W_+).
\end{equation}
Therefore the solutions will only depend on the functions $m(r)$, $S(r)$ and the complex components of $W_+(r)$.

The field equations in the case $\Lambda=0$ are well-known \cite{brodbeck_generalized_1993, oliynyk_local_2002}. It is not difficult to use the general adS Einstein and YM field equations to derive the analogous forms for $\Lambda<0$. These general field equations are also well-known:
\labeq{\begin{split}
		& 2T_{\mu\nu}=G_{\mu\nu}+\Lambda g_{\mu\nu},\\& 0=\nabla_\lambda F^{\,\,\lambda}_{\mu}+[A_\lambda,F^{\,\,\lambda}_\mu],\\
	\end{split}}{GenFEs}
	where $g_{\mu\nu}$ is the metric tensor defined using \eqref{metric}, $G_{\mu\nu}$ is the Einstein tensor, $F^{\,\,\lambda}_\mu$ is the mixed anti-symmetric field strength tensor defined with
	\labeq{F_{\mu\nu}=\partial_\mu A_\nu-\partial_\nu A_\mu + [A_\mu,A_\nu],}{Fmunu}
	$A_\mu$ represents the YM one-form connection \eqref{conn}, and the energy-momentum tensor $T_{\mu\nu}$ is given by
	\labeq{T_{\mu\nu}\equiv\mbox{Tr}\left[F_{\mu\lambda}F^{\,\,\lambda}_\nu-\frac{1}{4}g_{\mu\nu}F_{\lambda\sigma}F^{\lambda\sigma}\right].}{Tmunu}
	We note that Tr is the Lie algebra trace, we have used the Einstein summation convention where summation occurs over repeated indices, and we have rescaled all units so that 
	
	\labeq{4\pi G=c=q=1}{units}
	
	(for the gauge coupling constant $q$).
	
	Using \eqref{metric}, \eqref{conn} and \eqref{GenFEs}, we may show that the field equations for $\Lambda<0$ become
	\begin{subequations}\label{FEgen}
		\begin{align}
		\frac{dm}{dr}&=\mu G+\frac{P}{r^2},\label{EE1}\\
		\frac{1}{S}\frac{dS}{dr}&=\frac{2G}{r},\label{EE2}\\
		0&=r^2\mu W^{\prime\prime}_++2\left(m-\frac{P}{r}+\frac{r^3}{\ell^2}\right)W^\prime_++\mc{F},\label{YM}\\
		0&=[W_+,W^\prime_-]-[W^\prime_+,W_-],\label{YMB}
		\end{align}
	\end{subequations}
	with ${}^\prime\equiv d/dr$,
	\begin{equation}\label{quantdefs}
	\begin{array}{llll}
	\mu=1-\disfrac{2m}{r}+\disfrac{r^2}{\ell^2}, \quad & G\equiv\frac{1}{2}(W^\prime_+,W^\prime_-),\quad & \hat{F}\equiv -\frac{i}{2}\left(W_0-[W_+,W_-]\right), & \\
	&&&\\
	\mc{F}\equiv -i[\hat{F},W_+], & P\equiv-\frac{1}{2}(\hat{F},\hat{F}), & & \\
	\end{array}
	\end{equation}
	and $\ell$, the adS radius of curvature, given by
	\begin{equation}
	\ell\equiv\sqrt{\frac{-3}{\Lambda}},
	\end{equation}
	only valid for $\Lambda<0$. In \eqref{quantdefs}, $(\,\, , \,)$ is an invariant inner product on $\mk{g}$ determined up to a factor on each simple component of a semisimple $\mk{g}$ (arising from the Lie algebra trace), which induces a norm $|\,\,|$ on (the Euclidean) $\mk{h}$ and therefore also on its dual. These factors are chosen so that $(\,\, , \,)$ is a positive multiple of the Killing form on each simple component.
	
	We may calculate the energy density $e$, the radial pressure $p_r$ and the tangential pressure $p_\theta$. As we mentioned in Section \ref{sec:conn}, these are important quantities which help us assess the physicality of our solutions. First we note that since $c(\hat{F})=\hat{F}$, and $\HIP{X}{Y}\equiv-(c(X),Y)$ is a Hermitian inner product on $\mk{g}$, then $G\geq 0$ and $P\geq 0$. Then, we have (in our units \eqref{units})
	\begin{equation}
	e=r^{-2}(\mu G+r^{-2}P),\qquad p_r=r^{-2}(\mu G-r^{-2}P),\qquad p_\theta=r^{-4}P.
	\end{equation}
	
	Now we describe how to reduce the field equations down to the case of a regular action as described above. We select a Chevally-Weyl basis for $\mk{g}$. Let $R$ be the set of roots on $\mk{h}^*$ and $\Sigma=\{\alpha_1,...,\alpha_\mc{L}\}$ be a basis for $R$ (where $\mc{L}$ is the rank of $\mk{g}$). We also define
	\begin{equation}
	\langle\alpha,\beta\rangle\equiv\frac{2(\alpha,\beta)}{|\beta|^2},\qquad (\bv{t}_\alpha,X)\equiv\alpha(X)\quad\forall X\in\mk{h},\qquad \bv{h}_\alpha\equiv\frac{2\bv{t}_\alpha}{|\alpha|^2}.
	\end{equation}
	Then $\{\bv{h}_i\equiv\bv{h}_{\alpha_i},\bv{e}_\alpha,\bv{e}_{-\alpha}\,|\,i=1,...,\mc{L};\,\alpha\in R\}$ is a basis for $\mk{g}$, and induces the decomposition
	\begin{equation}
	\mk{g}=\mk{h}\oplus\bigoplus_{\alpha\in R^+}\mk{g}_\alpha\oplus\mk{g}_{-\alpha}
	\end{equation}
	for $R^+$, the set of positive roots expressed in the basis $\Sigma$. For this decomposition, we adopt the conventions
	\begin{equation}\label{basedef}
	[\bv{e}_{\alpha},\bv{e}_{-\alpha}]=\bv{h}_{\alpha},\qquad [\bv{e}_{-\alpha},\bv{e}_{-\beta}]=-[\bv{e}_{\alpha},\bv{e}_{\beta}],\qquad(\bv{e}_{\alpha},\bv{e}_{-\alpha})=\frac{2}{|\alpha|^2}.
	\end{equation}
	
	From the commutator relations defining an $\mk{sl}(2)$-subalgebra $\mbox{span}\{\bv{e}_{0},\bv{e}_{\pm}\}$ of $\mk{g}$, i.e.
	\begin{equation}
	[\bv{e}_{0},\bv{e}_{\pm}]=\pm2\bv{e}_{\pm},\qquad[\bv{e}_{+},\bv{e}_{-}]=\bv{e}_{0},
	\end{equation}
	and using
	\begin{equation}
	[\bv{h}_{},\bv{e}_{\alpha}]=\alpha(\bv{h}_{})\bv{e}_{\alpha},
	\end{equation}
	it follows \cite{dynkin_semisimple_1952} that $\bv{e}_{0}$ can only be an $A_1$-vector if there is an $\alpha\in R$ such that
	\begin{equation}
	\alpha(\bv{e}_0)=2. 
	\end{equation}
	Hence, writing $W_0$ in the basis
	\begin{equation}
	W_0=\slim_{i=1}^\mc{L}\lambda_i\bv{h}_{i}\in\mk{h},
	\end{equation}
	then equations \eqref{WE2} imply that
	\begin{equation}\label{Wplus}
	W_+(r)=\slim_{\alpha\in \Sigma_\lambda}\omega_\alpha(r)\bv{e}_{\alpha},
	\end{equation}
	where we have defined $\Sigma_\lambda$, a set of roots depending on the homomorphism $\lambda$ (or equivalently the constants $\lambda_i$), as
	\begin{equation}\label{slamdef}
	\Sigma_\lambda\equiv\{\alpha\in R\,|\,\alpha(W_0)=2\}.
	\end{equation}
	
	In a similar way we find that
	\begin{equation}
	W_-(r)=\slim_{\alpha\in \Sigma_\lambda}\varpi_\alpha(r)\bv{e}_{-\alpha},
	\end{equation}
	for functions $\varpi_\alpha(r)$, but given that the complex conjugation operator $c$ maps $\bv{h}_{i}\mapsto-\bv{h}_{i}$, $\bv{e}_{\alpha}\mapsto -\bv{e}_{-\alpha}$, we easily see that
	\begin{equation}
	\varpi_\alpha(r)=c(\omega_{\alpha}(r)).
	\end{equation}
	Therefore, the system is determined by two real functions $m(r)$, $S(r)$ and $\mc{L}$ complex functions $\omega_\alpha(r),\,\,\forall\alpha\in \Sigma_\lambda$.
	
	It is noted in \cite{oliynyk_local_2002} that we may na\"{i}vely proceed by substituting the expansion \eqref{Wplus} into the field equations and calculate the various Lie brackets using \eqref{basedef}, but this may produce many more equations that unknowns, and in addition there is still some gauge freedom left in the connection $\mc{A}$. However we may simplify the system a great deal by considering only the so-called $\textit{regular}$ case, where $W_0$ is a vector in the \emph{open} fundamental Weyl chamber $W(S)$ \cite{brodbeck_self-gravitating_1994}. We begin with a theorem due to Brodbeck and Straumann:
	\begin{thr} \cite{brodbeck_generalized_1993}\label{Kthm1}
		If $W_0$ is in the open Weyl chamber $W(\Sigma)$ then the set $\Sigma_\lambda$ is a $\Pi$-system, i.e. satisfies:
		
		\begin{enumerate}
			\item if $\alpha,\beta\in \Sigma_\lambda$ then $\alpha-\beta\notin R$,\\
			\item $\Sigma_\lambda$ is linearly independent;
		\end{enumerate}
		
		and is therefore the base of a root system $R_\lambda$ which generates a Lie subalgebra $\mk{g}_\lambda$ of $\mk{g}$ spanned by $\{\bv{\emph{h}}_{\alpha},\bv{\emph{e}}_{\alpha},\bv{\emph{e}}_{-\alpha}\,|\,\alpha\in R_\lambda\}$. Moreover, if $\mk{h}_\lambda\equiv\mbox{\emph{span}}\{\bv{\emph{h}}_{\alpha}\,|\,\alpha\in \Sigma_\lambda\}$ and $\mk{h}^\perp_\lambda\equiv\bigcap_{\alpha\in \Sigma_\lambda}\ker\alpha$ then
		\begin{equation}
		\mk{h}=\mk{h}_\lambda^\parallel\oplus\mk{h}^\perp_\lambda\quad\mbox{ and }\quad W_0=W_0^\parallel + W_0^\perp\quad\mbox{ with }\quad W_0^\parallel=\slim_{\alpha\in R_\lambda}\bv{\emph{h}}_{\alpha}.
		\end{equation}
		
		If $W_0$ is an $A_1$-vector then $W_0^\perp=0$ (though $\mk{h}_\lambda^\perp$ need not be trivial).
	\end{thr}
	This allows us to rewrite the field equations in a much simpler form -- in fact, in a form that renders them very similar-looking to the well-studied $\sun$ case.
	
	First we can consider $W_+$ to be a $\mk{g}_\lambda$-valued function, and write
	\begin{equation}
	W_{+}(r)=\slim_{j=1}^{\mc{L}_{\lambda}}\omega_j(r)\tilde{\bv{e}}_j,
	\end{equation}
	where we now take $\{\tilde{\alpha}_1,...,\tilde{\alpha}_{\mc{L}_\lambda}\}$ as the basis for $\Sigma_\lambda$ and define $\tilde{\bv{e}}_j\equiv\bv{e}_{\tilde{\alpha}_j}$. This means that using \eqref{basedef}, \eqref{YMB} becomes
	\begin{equation}\label{YM20}
	\slim_{j=1}^{\mc{L}_\lambda} \left(\omega_jc(\omega_j)^{\prime}-\omega^\prime_jc(\omega_j)\right)\bv{h}_{j}=0,
	\end{equation}
	implying that the phase of $\omega_j(r)$ is constant and can be set to zero using a gauge transformation. Hence we can conclude that the $\omega_j(r)$ may we taken as real-valued functions. We note that in $\Lambda=0$, this is only possible for the regular case \cite{oliynyk_local_2002}. Also using this basis, we may define the Cartan matrix of the reduced subalgebra $\mk{g}_\lambda$ as
	\begin{equation}
	C_{ij}\equiv\langle\tilde{\alpha}_i,\tilde{\alpha}_j\rangle,
	\end{equation}
	noting that by definition this is a symmetric and positive operator. 
	
	The results in Section 3 of \cite{oliynyk_local_2002} depend only on the root structure of the reduced subalgebra, and therefore we may also apply the same logic when reducing the field equations \eqref{FEgen} to the regular case. Finally then, dropping tildes from $\alpha_j$ and losing the $\lambda$ index from $\mk{g}$ \textit{et cetera} for clarity, we can show that the field equations become
	\begin{subequations}\label{NewFEs}
		\begin{align}
		m'&=\mu G+\frac{P}{r^2},\label{EE12}\\
		\frac{S'}{S}&=\frac{2G}{r},\label{EE22}\\
		0&=r^2\mu\omega_j^{\prime\prime}+2\left(m-\frac{P}{r}+\frac{r^3}{\ell^{2}}\right)\omega^\prime_j+\frac{1}{2}\slim_{k=1}^\mc{L}\omega_jC_{jk}(\lambda_k-\omega^2_k),\label{YM2}
		\end{align}
	\end{subequations}
	with
	\begin{subequations}\label{NewQuants}
		\begin{align}
		\mu&=1-\frac{2m}{r}+\frac{r^2}{\ell^2},\\
		P&=\frac{1}{8}\slim_{j,k=1}^\mc{L}(\lambda_j-\omega_j^2)h_{jk}(\lambda_k-\omega_k^2)\label{Pdef2},\\
		G&=\slim_{k=1}^\mc{L}\frac{\omega_k^{\prime 2}}{|\alpha_k|^2},\label{Gdef2}\\
		h_{jk}&=\frac{2C_{jk}}{|\alpha_j|^2}.\label{hdef}
		\end{align}
	\end{subequations}
	
	The final step is to determine the values of the constants $\lambda_j$, which involves determining the subalgebra $\mk{g}_\lambda$ for a given $A_1$-vector $W_0$ in the open fundamental Weyl chamber. For a semisimple group, for which the Cartan subalgebra splits into an orthogonal sum $\mk{h}=\bigoplus_k\mk{h}_k$, the orthogonal decomposition given in Theorem \ref{Kthm1} splits into analogous decompositions of each of $\mk{h}_k$. Hence we only need consider the regular actions of simple Lie groups.
	
	However, we note that the $A_1$-vector in the Cartan subalgebra $\mk{h}$ of a Lie algebra $\mk{g}$ is uniquely determined by the integers
	\begin{equation}\label{chidef}
	\{\chi_1,...,\chi_\mc{L}\}\equiv\{\alpha_1(W_0),...,\alpha_\mc{L}(W_0)\},
	\end{equation}
	
	which integers are chosen from the set $\{0,1,2\}$. In \cite{dynkin_semisimple_1952}, this is referred to as the \emph{characteristic}. From \eqref{slamdef}, it is obvious that for the principal action, 
	\begin{equation}\label{chidefprin}
	\chi_j=2\quad(\forall j\in\{1,...,\mc{L}\})
	\end{equation}
	for $\mk{h}_\lambda$. $A_1$-vectors satisfying this define \emph{principal $\essu$-subalgebras}, and hence \emph{principal actions} of $SU(2)$ on the bundle. As in \cite{oliynyk_local_2002}, we may rely the following theorem:
	
	\begin{thr}\cite{oliynyk_local_2002}\label{Kthm2}
		\begin{enumerate}
			\item The possible regular $\essu$-subalgebras of simple Lie algebras consist of the principal subalgebras of all Lie algebras $A_\mc{L}$, $B_\mc{L}$, $C_\mc{L}$, $D_\mc{L}$, $G_2$, $F_4$, $E_6$, $E_7$ and $E_8$ and of those subalgebras  of $A_\mc{L}=\mk{sl}(\mc{L}+1)$ with even $\mc{L}$ corresponding to partitions $[\mc{L}+1-k,k]$ for any integer $k=1,...,\mc{L}/2$, or, equivalently, characteristic (22...2211...1122...22) ($2k$ `1's in the middle and `2's in all other positions);\\
			\item The Lie algebra $\mk{g}_\lambda$ is equal to $\mk{g}$ in the principal case, and for $A_\mc{L}$ with even $\mc{L}$ equal to $A_{\mc{L}-1}$ for $k=1$ and to $A_{\mc{L}-k}\oplus A_{k-1}$ for $k=2,...,\mc{L}/2$;\\
			\item In the principal case $\mk{h}_\lambda^\parallel=\mk{h}$. For all $\essu$-subalgebras of $A_\mc{L}$ with even $\mc{L}$ the orthogonal space $\mk{h}_\lambda^\perp$ is one-dimensional.
		\end{enumerate}
	\end{thr}
	
	The essence of this theorem is that the regular action here coincides with the principal action. This finally allows us to determine an expression for the constants $\lambda_j$, derived by using \eqref{Pdef2}, \eqref{hdef}, \eqref{chidefprin}, and \eqref{lamdef}:
	\begin{equation}
	\lambda_j=2\slim_{k=1}^\mc{L}(C^{-1})_{jk}.\label{lamdef}
	\end{equation}

\section{Boundary conditions}\label{sec:BCs}

In order to get a sense of the possible term dependencies in the power series expansions of the field variables near the boundary points, and thus decide what methods we will need to prove local existence, it is very enlightening to calculate the lower order terms in the power series expansions of the field variables nearby the boundaries $r=0$, $r=r_h$ and $r\rar\infty$. We do this below, in anticipation of the later proofs of local existence at these points in Section \ref{sec:locex}.

In the black hole case, i.e. for the boundaries $r=r_h$ and $r\rar\infty$, we find that the situation is relatively uncomplicated. For $r=r_h$, the lower order terms show that the solutions can be characterised entirely by the values of $\omega_j(r_h)\equiv\omega_{j,h},\,\,\,\forall j=1,...,\mc{L}$. Asymptotically, we find that the solution is parametrised entirely by the values of the limits of $m(r)$, $\omega_j(r)$ and $r^2\omega^\prime_j(r)$ ($j=1,...,\mc{L}$) as $r\rar\infty$. We find no constraints on the boundary values of the field variables asymptotically, and near $r=r_h$, we merely find a couple of constraints on the metric function $\mu(r)$ that must be satisfied, which are physically necessary to ensure a regular and non-extremal event horizon.

In the soliton case however, i.e. at $r=0$, the situation is much more complicated, as it was in the $\sun$ case \cite{baxter_existence_2008, baxter_existence_2016}. There, we had to solve a tridiagonal matrix equation by using expansions in the eigenvectors of the matrix in question; for this we used Hahn polynomials, an orthogonal class of polynomials defined using hypergeometric functions \cite{samuel_karlin_hahn_1961}. In that case, as in this, $\Lambda$ appears at $O(r^2)$ and above in the field equations \eqref{EE1} -- \eqref{YM}, and therefore near $r=0$ we do not expect the appearance of the cosmological constant to make any appreciable difference. 

In light of all of this, we now review the boundary conditions we expect in each case.

\subsection{Origin}

Near $r=0$ we may simply use the independent variable $r$, and hence we expand all field variables and quantities as
\begin{equation}
f(r)=\slim_{k=0}^\infty f_kr^k
\end{equation}

for a general function $f(r)$. Thus we obtain the following recurrence relations for $m_{k+1}$, $S_{k}$ and $\omega_{j,k+1}$:
\begin{subequations}\label{recur0}
	\begin{align}
	(k+1)m_{k+1}&=G_k+\frac{1}{\ell^2}G_{k-2}+P_{k+2}-2\slim_{l=2}^{k-2}m_{k-l}G_l,\label{rrorm}\\
	kS_k&=2G_k,\label{rrors}\\
	b_{i,k}&=\slim_{j=1}^\mc{L}\left(A_{ij}-k(k+1)\bsym{\delta}_{ij}\right)\omega_{j,k+1}.\label{rrorb}
	\end{align}
\end{subequations}

Here, $\bv{A}\equiv A_{ij}$ is the matrix defined by
\labeq{A_{ij}\equiv\omega_{i,0}C_{ij}\omega_{j,0}\quad\mbox{(no sum on }i,j);}{Adef}
$\bsym{\delta}_{ij}$ is the Kronecker symbol; and the left-hand side of \eqref{rrorb}, the vector $\bv{b}_k\equiv(b_{1,k},...,b_{\mc{L},k})$, is a complicated vector expression involving the coefficients of the field variable expansions. 

We can see that these equations are identical to the $\sun$ case \cite{kunzle_analysis_1994}, and so again, we may solve \eqref{rrorm} and \eqref{rrors} and obtain a solution with $\mc{L}$ free parameters on condition that the recurrence relations \eqref{rrorb} can be solved. This in turn is conditional upon the vectors $\bv{b}_k$ lying in the left kernel of the matrix $\bv{A}$. As we noted, $\bv{b}_k$ is a complicated expression and so this is difficult to prove in general. In Section \ref{sec:lex0}, we generalise proofs in \cite{oliynyk_local_2002} which depend directly on the root structure of the Lie algebra $\mk{g}$ treated as an $\mk{sl}(2,\C)$ submodule.

We note here that $G_k=P_k=0$ for $k<2$. For the lower order terms, we find:
\begin{equation}
S_0\neq 0,\quad m_0=m_1=m_2=0,\quad \omega^2_{j,0}=\lambda_j,\quad \omega_{j,1}=0.
\end{equation}

The equations \eqref{recur0} are identical to those we found in the $\sun$ case, therefore we expect a similar situation to occur here, in that the higher order terms of the power series expansions near the origin will in general display a complicated interdependence. This reflects the fact that $r=0$ is a singular point of the field equations. At this boundary, the higher order coefficients which remain arbitrary occur at the orders $r^k$ for which $k(k+1)$ is an eigenvalue of the matrix $\bv{A}$. But in fact, the eigenvalues of $\bv{A}$ can happily be shown to be $k(k+1)$ for a series of integer values of $k$, which series depends on the Lie algebra in question. (For $\sun$, this series of integers is simply the natural numbers from 1 to $N-1$ inclusive.) For all the simple Lie algebras, we may calculate the spectrum of eigenvalues from the Cartan matrix by using the definition \eqref{Adef} -- see Table \ref{Table1} for this information. The proof for the classical Lie algebras then follows from the properties of the root structure and the results at the end of Section \ref{sec:sl2c}.

We will see in Section \ref{sec_lex0} that in some neighbourhood of $r=0$, the relevant field variables have the following behaviour:
\begin{equation}
\begin{split}
m(r)&=m_3r^3+O(r^4),\\
S(r)&=S_0+O(r^2),\\
\omega_i(r)&=\omega_{i,0}+\slim_{j=1}^\mc{L} Q_{ij}\hat{u}_j(r)r^{k_j+1},\quad i=1,...,\mc{L}.
\end{split}
\end{equation}

Here, $Q_{ij}$ is a non-singular matrix, $k_j$ are integers and $\hat{u}_j$ are some functions of $r$ -- all of these we will define later. Also, $m_3$ is fixed by \eqref{rrorm}, $S_0$ is fixed by the requirement that $S\rar1$ as $r\rar\infty$, and $\omega^2_{j,0}=\lambda_j$. Therefore altogether we have $\mc{L}$ free solution parameters here in total, namely $\hat{u}_j(0)$ for each $j$.

\begin{table}
	\centering
	\begin{tabular}{ || c | c || }
		\hline\hline
		Lie algebra & $\mc{E}$\\
		\hline
		\hline
		Classical: & \\
		\hline
		$A_\mc{L}$ & $j$\\
		\hline
		$B_\mc{L}$ & $2j-1$\\
		\hline
		$C_\mc{L}$ & $2j-1$\\
		\hline
		$D_\mc{L}$ & 
		$\Bigg\{\begin{array}{ll}
		2j-1 & \mbox{ if }j\leq\,(\mc{L}+2)/2\\
		\mc{L}-1 & \mbox{ if }j=(\mc{L}+2)/2\\
		2j-3 & \mbox{ if }j\mbox{ \textgreater }\,(\mc{L}+2)/2\\
		\end{array}$\\
		\hline
		Exceptional: & \\
		\hline
		$G_2$ & 1, 5\\
		\hline
		$F_4$ & 1, 5, 7, 11\\
		\hline
		$E_6$ & 1, 4, 5, 7, 8, 11 \\
		\hline
		$E_7$ & 1, 5, 7, 9, 11, 13, 17\\
		\hline
		$E_8$ & 1, 7, 11, 13, 17, 19, 23, 29\\
		\hline\hline
	\end{tabular}
	\caption{This table shows $\mbox{spec}(\bv{A})=\{k(k+1)\,|\,k\in\mc{E}\}$. For the classical Lie algebras the table shows $k_j$ for $j=1,...,\mc{L}$, $\mc{L}=\mbox{rank}(\mk{g})$. Note that $k=1$ belongs to all Lie algebras, thus $1\in\mc{E}$ always.}
	\label{Table1}
	\centering
\end{table}

\subsection{Event horizon}\label{ssec:bcsrh}

For a regular non-extremal event horizon, we require $\mu_h$ to vanish and $\mu'_h$ to be finite and positive. This severely restricts the solution parameters here and hence reduces the degrees of freedom of any solution, which makes boundary conditions easy to find.

Using the notation $f_h\equiv f(r_h)$ and transforming to a new variable $\rho=r-r_h$, we find that
\begin{equation}
\begin{split}
\mu(\rho)&=\mu^\prime_h\rho+O(\rho^2),\\
S(r)&=S_h+O(\rho),\\
\omega_j(\rho)&=\omega_{j,h}+O(\rho),\\
\end{split}
\end{equation}
where
\begin{equation}
\mu^\prime_h=\frac{1}{r_h}+\frac{3r_h}{\ell^2}-\frac{2}{r_h^3}P_h.
\end{equation}

The constraint $\mu_h=0$ implies that
\begin{equation}
\begin{split}
m_h&=\frac{r_h}{2}+\frac{r^3_h}{2\ell^2},\quad\mbox{and}\\
\omega^\prime_{j,h}&=-\frac{\mc{F}_{j,h}}{2\left(m_h-r_h^{-1}P_h+r_h^3\ell^{-2}\right)},\\
\end{split}
\end{equation}
with
\begin{equation}
\mc{F}_{j,h}=\frac{1}{2}\omega_{j,h}\slim_{k=1}^\mc{L}C_{jk}(\lambda_k-\omega_{k,h}^2).
\end{equation}

The condition $\mu^\prime_h>0$ places a bound on $m^\prime_h$:

\begin{equation}
m^\prime_h=\frac{P_h}{r_h^2}>0,
\end{equation}
with
\begin{equation}
P_h=\frac{1}{8}\slim_{j,k=1}^\mc{L}(\lambda_j-\omega_{j,h}^2)h_{jk}(\lambda_k-\omega_{k,h}^2).
\end{equation}
Therefore, it is clear that fixing $r_h$ and $\ell$, and regarding $S_h$ as fixed by the requirement that the solution is asymptotically adS, the solution parameters are given by the set $\{\omega_{j,h}\}$. Thus, as at the origin, we have $\mc{L}$ solution degrees of freedom for solutions existing locally at the event horizon. 

\subsection{Infinity}

We assume power series for all field variables which are good in the asymptotic limit, i.e. of the form $f(r)=f_\infty+f_1r^{-1}+...$. It is easy to see that this implies $G=O(r^{-4})$, meaning that examining \eqref{EE2}, $S$ must be of the form $S(r)=S_\infty+O(r^{-4})$. We also use the basis $W_+(r)=\sum_{j=1}^\mc{L}\omega_j(r)\bv{e}_{\alpha_j}$. Therefore, we find that the expansions near infinity must be
\begin{equation}\label{BCinf}
\begin{split}
m(r)&=m_\infty+m_1r^{-1}+O(r^{-2}),\\
S(r)&=S_\infty+S_4r^{-4}+O(r^{-5}),\\
\omega_j(r)&=\omega_{j,\infty}+c_jr^{-1}+d_jr^{-2}+O(r^{-3}).\\
\end{split}
\end{equation}

The power series expansions here are a lot less complicated than for the asymptotically flat case. No constraints appear on $\omega_{j,\infty}$ or $c_j$. Similarly, no constraints are placed on $S_\infty$ or $m_\infty$, so we rescale to $S_\infty=1$ and let $m_\infty=M$ (the constant Arnowitt-Deser-Misner (ADM) mass) so that the solution asymptotically is the SadS solution (or pure adS space if $M=0$). We find that each new term we calculate in the expansions is entirely determined by previously calculated terms, and this trend continues for higher order terms. For instance, the lower order terms are
\begin{equation}
\begin{split}
m_1&=-\frac{1}{\ell^2}\slim_{j=1}^{\mc{L}}\frac{c_j^2}{|\alpha_j|^2}-\slim_{j,k=1}^{\mc{L}}(\lambda_j-\omega_{j,\infty}^2)h_{jk}(\lambda_k-\omega_{k,\infty}^2),\\
S_4&=-\frac{1}{2}\slim_{j=1}^{\mc{L}}\frac{c_j^2}{|\alpha_j|^2},\\
d_j&=\frac{\ell^2}{4}\omega_{j,\infty}\slim_{k=1}^{\mc{L}}C_{jk}(\lambda_k-\omega_{k,\infty}^2).\\
\end{split}
\end{equation}
Therefore we anticipate that proving the existence of unique solutions to the boundary value problem will be a lot less involved than in the case of $\Lambda=0$. In summary, our solution parameters here are $\{M,\omega_{j,h},c_j\}$ and thus we have $2\mc{L}+1$ degrees of freedom in total.

\section{Asymptotic behaviour of the field equations}\label{sec:asym}

As we saw, the asymptotic boundary conditions \eqref{BCinf} imply that any regular solutions in this limit will have gauge functions which are characterised entirely by the arbitrary values $\omega_{j,\infty}$ and $c_j$, with all higher order terms in the expansions determined by these parameters. This is in opposition to the $\Lambda=0$ case, where the asymptotic values of the gauge field have to approach particular values, and the higher order terms display complicated interdependence related to the intercoupling of the gauge functions caused by equation \eqref{rrorb}.

Therefore what we wish to do now is take the asymptotic limit of the field equations, transform the independent variable $r$ so that the system becomes `autonomous' in the dynamical systems sense, and examine the nature of the phase plane of the system. As we will see, it is not so much the asymptotic field equations themselves which give us the difference in behaviour between the $\Lambda=0$ and $\Lambda<0$ cases -- it is the form of the parameter we must transform to which dictates the asymptotic behaviour of the field variables, and which gives us an infinitely more plentiful space of regular solutions.

First, we note that as $r\rar\infty$, $\mu\approx 1+\frac{r^2}{\ell^2}$. Noting also \eqref{BCinf}, the YM field equations \eqref{YM} become asymptotically
\begin{equation}\label{YMasym}
\frac{r^4}{\ell^2}W^{\prime\prime}_++\frac{2r^3}{\ell^2}W^\prime_++\mc{F}=0.
\end{equation}
Using the parameter $\tau=\ell r^{-1}$, we find that \eqref{YMasym} becomes
\begin{equation}\label{Wplusasym}
\frac{d^2W_+}{d\tau^2}=-\mc{F}.
\end{equation}
In the more explicit basis \eqref{Wplus} defined in Section \ref{sec:FEs}, where the field equations become \eqref{NewFEs}, this is equivalent to
\begin{equation}
\frac{d^2\omega_j}{d\tau^2}=-\frac{1}{2}\slim_{k=1}^{\mc{L}}\omega_jC_{jk}(\lambda_k-\omega_{k}^2).
\end{equation}

It is easy to see that the critical points $\omega^*_j$ of this autonomous system satisfy $\mc{F}=0$, i.e. where
\begin{equation}
\omega^*_j\slim_{k=1}^{\mc{L}}C_{jk}(\lambda_k-\omega_k^{*2})=0.
\end{equation}
Noting that $C_{ij}$ is of full rank, this gives us two sets of critical points: either $\omega^*_j=0$, or $\omega^{*}_j=\pm\lambda^{1/2}_j$, $\forall j\in\{1,...,\mc{L}\}$. Eigenvalue analysis shows these (for each $j$) to be a centre and a pair of saddles, respectively. We noted that the analysis of the asymptotic boundary conditions \eqref{BCinf} implied no such constraints on the asymptotic value of $\omega_j(r)$, though the autonomous asymptotic equations \eqref{Wplusasym} are identical to those for $\Lambda=0$.

We may resolve this apparent discrepancy by noting that for $\Lambda<0$, the trajectory of a solution in the phase plane $\left(\omega_j,\frac{d\omega_j}{d\tau}\right)$ will not in general reach its critical point. This is due to the nature of the parameter we used to render the equations autonomous. In the case of $\Lambda=0$ the parameter used was $\tau\propto\log r$, so that the range $r\in[r_0,\infty)$ ($r_0=r_h$ for black holes, or $r_0=0$ for solitons) corresponds to $\tau\in(-\infty,\infty)$, and hence any trajectory for a regular solution in the limit $r\rar\infty$ will be destined to end at a critical point. 

For $\Lambda<0$ however, we use $\tau\propto 1/r$, meaning that the range $r\in[r_0,\infty)$ corresponds to the range $\tau\in[0,r_0^{-1})$. Therefore, as we take the asymptotic limit $r\rar\infty$, the corresponding trajectories in terms of $\tau$ will shrink and only traverse a short distance in the phase plane. Hence the trajectories, and therefore the values of the gauge field functions and their derivatives, will in general approach arbitrary values asymptotically. We note that this is precisely the same as in the $\sun$ case \cite{baxter_existence_2008}. 

In summary then, our investigation has shown that we need not be concerned with the behaviour of the field equations for $r$ arbitrarily large -- as long as we can integrate into the asymptotic region, the solution will remain regular until reaching the (arbitrary) boundary conditions at $r\rar\infty$. We will return to this point in Section \ref{sec:gloex}.

\section{Embedded solutions}\label{sec:embed}

Our argument in Section \ref{sec:gloex} will rely on the existence of embedded (or `trivial') solutions, as we will prove the existence of global solutions to the field equations \eqref{EE12} to \eqref{YM2} in some neighbourhood of these. Therefore, we here review some easily obtainable embedded solutions to our field equations.

\subsection{Reissner-N\"{o}rdstrom anti-de Sitter (RNadS)}\label{sec:RNadS}

Here we let $\omega_j(r)\equiv0$. In that case, we find that $G=\mc{F}=0$ and therefore $S$ becomes a constant, which we scale to 1. The metric function $\mu(r)$ becomes
\begin{equation}
\mu=1-\frac{2M}{r}+\frac{Q^2}{r^2}+\frac{r^2}{\ell^2},
\end{equation}
where $M$ is the ADM mass of the solution, and the magnetic charge $Q$ is defined with
\begin{equation}\label{Qdef}
Q^2\equiv 2P\big|_{\omega_j\equiv 0}=\frac{1}{4}\slim_{j,k=1}^\mc{L}\lambda_jh_{jk}\lambda_k.
\end{equation}

Therefore we have obtained the embedded Reissner-N\"{o}rdstrom anti-de Sitter solution, which only exists with this value of $Q^2$, and coincides with the $\sun$ case \cite{baxter_existence_2008}, using \eqref{lamdef} and the $\sun$ Cartan matrix.

To summarise, the RNadS solution is given by
\begin{equation}
m(r)\equiv M,\qquad S(r)\equiv 1,\qquad\omega_j(r)\equiv 0,\qquad\forall r,\forall j=1,...,\mc{L}.
\end{equation}

\subsection{Schwarzschild anti-de Sitter (SadS)}\label{sec:SadS}

Here we let $\omega^2_j(r)\equiv\lambda_j,\,\,\,\forall r,\forall j=1,...,\mc{L}$. Then from \eqref{NewQuants} we find that $P=G=\mc{F}=0$, implying the following. From \eqref{EE12}, we get $m^\prime(r)=0$, so that $m(r)$ is a constant which we again set to the ADM mass $M$. From \eqref{EE22} we have $S^\prime(r)=0$, so that $S$ is a constant which we scale to $1$ for the asymptotic limit. Finally, the YM equations \eqref{YM2} are automatically satisfied. Since $P=0$, this solution carries no global charge, and can be identified as the embedded Schwarzschild anti-de Sitter solution. Summarising this solution:
\begin{equation}
m(r)\equiv M,\qquad S(r)\equiv 1,\qquad\omega^2_j(r)\equiv\lambda_j,\qquad\forall r,\forall j=1,...,\mc{L}.
\end{equation}

\subsection{Embedded $\essu$ solutions}\label{sec:su2embed}

Noting that we can embed $SU(2)$ isomorphically into any semisimple gauge group $G$, then there must always exist trivial embedded $\essu$ solutions to the field equations \eqref{EE1} to \eqref{YM} . We may show this by a simple rescaling.

\begin{prop}\label{prop:su2embed}
	Any solution to the field equations \eqref{EE1} -- \eqref{YM} can be rescaled and embedded as a solution which satisfies the field equations for $\essu$ adS EYM theory.
\end{prop}

\textbf{Proof} Consider the gauge group $G$, fixing the symmetry action such that $W_0$ is regular. Select any basis such that the set $\{W_0,\Omega_+,\Omega_-\}$ spans $\essu$, with $c(\Omega_+)=-\Omega_-$. We rescale the field variables as follows:
\begin{equation}
r=Q^{-1}\bar{r},\qquad \omega_j(r)\equiv \lambda_j\omega(\bar{r}),\qquad m\equiv Q\tilde{m}(\bar{r}),\qquad \ell\equiv Q\tilde{\ell},
\end{equation}
with $Q^2$ given in \eqref{Qdef}. Then the field equations \eqref{EE1} -- \eqref{YM} become
\begin{equation}
\begin{split}
\frac{d\tilde{m}}{d\bar{r}}&=\mu\left(\frac{d\omega}{d\bar{r}}\right)^2+\frac{(1-\omega^2)^2}{2\bar{r}^2},\\
\frac{1}{S}\frac{dS}{d\bar{r}}&=\frac{2}{\bar{r}}\left(\frac{d\omega}{d\bar{r}}\right)^2,\\
0&=\bar{r}^2\mu\frac{d^2\omega}{d\bar{r}^2}+\left(2\tilde{m}-\frac{(1-\omega^2)^2}{\bar{r}}+\frac{\bar{r}^3}{\tilde{\ell}^2}\right)\frac{d\omega}{d\bar{r}}+\omega(1-\omega^2),
\end{split}
\end{equation}
with
\labeq{\mu(\bar{r})=1-\frac{2\tilde{m}}{\bar{r}}+\frac{\bar{r}^2}{\tilde{\ell}^2}.}{}
These equations are identical to those for the $\essu$ adS case, for which the existence of (nodeless) solutions has been proven \cite{winstanley_existence_1999}. $\Box$

It is interesting to note that the scaling involves the magnetic charge itself, which can possibly be put down to the fact that the RNadS solution for $\essu$, embedded in the $\essu$ equations, only exists where the magnetic charge $Q^2=1$.

\section{Local existence proofs at the boundaries}\label{sec:locex}

Now we have much information about the behaviour of the solutions to the field equations nearby the boundaries of our spacetime, enough to prove local existence at those boundaries. To do this, we rely on a well-known theorem of differential equations \cite{breitenlohner_static_1994}, generalised to the appropriate case by \cite{oliynyk_local_2002}.

\begin{thr} \cite{oliynyk_local_2002}\label{Kthm3}
	The system of differential equations
	\begin{equation}
	\begin{split}
	t\frac{du_i}{dt}&=t^{\mu_i}f_i(t,u,v),\\
	t\frac{dv_i}{dt}&=-h_j(u)v_j+t^{\nu_j}g_j(t,u,v),\\
	\end{split}
	\end{equation}
	where $\mu_i,\,\nu_j\in\Z_{>1}$, $f_i$, $g_j$ are analytic functions in a neighbourhood of $(0,c_0,0)\in\R^{1+m+n}$, and the functions $h_j\,:\,\R^m\rar\R$ are positive in a neighbourhood of $c_0\in\R^m$, has a unique solution $t\mapsto(u_i(t),v_j(t))$ such that
	\begin{equation}
	u_i(t)=c_i+O(t^{\mu_i}),\qquad\mbox{ and }v_j(t)=O(t^{\nu_i}),
	\end{equation}
	for $|t|>\bar{r}$ for some $\bar{r}>0$ if $|c-c_0|$ is small enough. Moreover, the solution depends analytically on the parameters $c_i$.
\end{thr}

Essentially, the proof of this theorem proceeds from the requirement that formal power series may be found for the field variables at the boundaries in question. We now consider those boundaries one by one.

\subsection{Existence at the origin: $r=0$}\label{sec:lex0}

As we hinted in Section \ref{sec:BCs}, we do not expect much of a difference between the asymptotically flat and asymptotically adS cases nearby the origin, because as $r\rar0$, the terms in the field equations involving the cosmological constant become negligible. Hence we may proceed along very similar lines to those in \cite{oliynyk_local_2002}.

Therefore, we now collect all necessary results from \cite{oliynyk_local_2002} needed to prove local existence of solutions near $r=0$. The general idea is to consider the root structure of $\mk{sl}(2,\mathbb{C})$ taken as a Lie algebra submodule of $\mk{g}$. Note that the results in this Section are only necessary for this boundary, and hence only for solitons.

\subsubsection{Necessary results for local existence at $r=0$}\label{sec:sl2c}

First we introduce our conventions. We begin by defining a non-degenerate Hermitian inner product $\HIP{\,}{\,}:\mk{g}\times\mk{g}\rar\C$, such that
\begin{equation}
\langle X\,|\,Y\rangle\equiv-(c(X),Y)\quad\forall\,X,Y\in\mk{g}.
\end{equation}
Then $\HIP{\,}{\,}$ is a real positive definite inner product on $\mk{g}_0$, since $c:\mk{g}\rar\mk{g}$ is the conjugation operator determined on the compact real form $\mk{g}_0$. It is elementary to show that $\HIP{\,}{\,}$ satisfies
\begin{equation}
\begin{split}
\langle X\,|\,Y\rangle&=\overline{\langle Y\,|\,X\rangle},\\
\HIP{c(X)}{c(Y)}&=\overline{\langle X\,|\,Y\rangle},\\
\HIP{[X,c(Y)]}{Z}&=\HIP{X}{[Y,Z]}
\end{split}
\end{equation}
for all $X,Y,Z\in\mk{g}$. Now we introduce a positive definite, real inner product $\HHIP{\,}{\,}:\mk{g}\times\mk{g}\rar\R$, with
\begin{equation}\label{HHIP}
\HHIP{X}{Y}\equiv\mbox{Re}\HIP{X}{Y}\quad\forall X,Y\in\mk{g}.
\end{equation}
Let $\|\,\,\|$ be the norm induced by \eqref{HHIP}, i.e. $\|X\|^2=\HHIP{X}{X}\,\forall X\in\mk{g}$. Then we can easily verify the following properties of $\HHIP{\,}{\,}$:
\begin{equation}
\begin{split}
\HHIP{X}{Y}&=\HHIP{Y}{X},\\
\HHIP{c(X)}{c(Y)}&=\HHIP{X}{Y},\\
\HHIP{[X,c(Y)]}{Z}&=\HHIP{X}{[Y,Z]}
\end{split}
\end{equation}
for all $X,Y,Z\in\mk{g}$. 

Let $\Omega_+,\Omega_-\in\mk{g}$ be two vectors such that
\begin{equation}
[W_0,\Omega_{\pm}]=\pm 2\Omega_\pm,\quad\quad [\Omega_+,\Omega_-]=W_0,\quad\quad c(\Omega_+)=-\Omega_-.
\end{equation}
Then $\mbox{span}_\C\{W_0,\Omega_+,\Omega_-\}\cong\mk{sl}(2,\C)$. We again use a central dot notation $\cdot$ to represent the adjoint action, i.e.
\begin{equation}
X\ccdot Y\equiv\mbox{ad}(X)(Y),\qquad\forall X\in\mbox{span}_\C\{W_0,\Omega_+,\Omega_-\},\,Y\in\mk{g}.
\end{equation}
But since $W_0$ is a semisimple element, $\mbox{ad}(W_0)$ is diagonalisable, and so from $\mk{sl}(2)$ representation theory we know that the eigenvalues are integers. Therefore we define $V_n$ as the eigenspaces of $\mbox{ad}(W_0)$, i.e. with
\begin{equation}
V_n\equiv\{X\in\mk{g}\,|\,W_0\ccdot X=nX,\,n\in\Z\,\}.
\end{equation}
It also follows from $\mk{sl}(2,\C)$ representation theory that if $X\in\mk{g}$ is a highest weight vector of the adjoint representation of $\mbox{span}_\C\{W_0,\Omega_+,\Omega_-\}$ with weight $n$, and we define $X_{-1}=0$, $X_0=X$ and $X_j=(1/j!)\Omega_-^j\ccdot X_0$ ($j\geq 0$), then
\begin{equation}
\begin{split}
W_0\ccdot X_j & =(n-2)X_j,\\
\Omega_-\ccdot X_j & =(j+1)X_{j+1},\\
\Omega_+\ccdot X_j & =(n-j+1)X_{j-1}.
\end{split}
\end{equation}

Now we are ready to state a series of results proven in \cite{oliynyk_local_2002} which will help us to prove existence locally at $r=0$. Essentially, these are necessary because we find that the term $\mc{F}$ in the YM equation \eqref{YM} is the only term which resists our rearrangement of the field equations in a form appropriate to Theorem \ref{Kthm3}, and it is necessary to argue that certain lower order term of $\mc{F}$ (in a power series sense) are zero. Hence we proceed.

\begin{prop}\label{Kprop1}
	There exist $\bsym{\Sigma}$ highest weight vectors $\xi^1$, $\xi^2,$,... $\xi^{\bsym{\Sigma}}$ for the adjoint representation of  $\emph{span}_\C\{W_0,\Omega_+,\Omega_-\}$ on $\mk{g}$ that satisfy
	\renewcommand{\labelenumii}{\Roman{enumii}}
	\begin{enumerate}
		\item the $\xi^j$ have weights $2k_j$ where $j=1,...,\bsym{\Sigma}$ and $1=k_1\leq k_2\leq...\leq k_{\bsym{\Sigma}}$;\\
		\item if $V(\xi^j)$ denotes the irreducible submodule of $\mk{g}$ generated by $\xi^j$, then the sum $\slim_{j=1}^{\bsym{\Sigma}}V(\xi^j)$ is direct;\\
		\item if $\xi^j_l=(1/l!)\Omega_-^l\ccdot\xi^j$, then $c(\xi^j_l)=(-1)^l\xi^j_{2k_j-l}$;\\
		\item $\bsym{\Sigma}=|\Sigma_\lambda|$ and the set $\{\xi^j_{k_j-1}\,|\,j=1,...,\bsym{\Sigma}\}$ forms a basis for $V_2$ over $\C$.\\
		
	\end{enumerate}
\end{prop}

\begin{prop}\label{Kprop2}
	The R-linear operator $A:\mk{g}\rar\mk{g}$ defined by
	\begin{equation}
	A\equiv\frac{1}{2}\emph{ad}(\Omega_+)\circ\left(\emph{ad}(\Omega_-)+\emph{ad}(\Omega_+)\circ c\right),
	\end{equation}
	is symmetric with respect to the inner product $\HHIP{\,}{\,}$, i.e. $\HHIP{A(X)}{Y}=\HHIP{X}{A(Y)}\,\,\forall X,Y\in\mk{g}$.
\end{prop}

\begin{lem}\label{Klem1}
	\begin{equation}
	A(V_2)\subset V_2.
	\end{equation}
\end{lem}

This shows that the operator $A$ restricts to $V_2$: we therefore denote this operator by
\begin{equation}
A_2\equiv A|_{V_2}.
\end{equation}

Now we label the set of integers $k_j$ from Proposition \ref{Kprop1} as follows:
\begin{equation}
\begin{split}
1=k_{J_1}=k_{J_1+1}=...=k_{J_1+k_1-1}&< k_{J_2}=k_{J_2+1}=...=k_{J_2+m_2-1}\\
&<...\\
&<k_{J_I}=k_{J_I+1}=...=k_{J_I+m_I-1},\\
\end{split}
\end{equation}
where we define the series of integers $J_1=1$, $J_k+m_k=J_{k+1}$ for $k=1,...,I$ and $J_{I+1}=\bsym{\Sigma}-1$. To ease notation we define
\begin{equation}\label{kapdef}
\kappa_j\equiv k_{J_j},\mbox{   for }j=1,...,I.
\end{equation}

As noted in Proposition \ref{Kprop1}, the set $\{\xi^j_{k_j-1}\,|\,j=1,...,\bsym{\Sigma}\}$ forms a basis of $V_2$ over $\C$. Therefore the set of vectors $\{X^l_s,Y^l_s\,|\,l=1,...,I;\,s=0,1,...,m_l-1\}$ forms a basis of $V_2$ over $\R$, where
\begin{equation}
X^l_s\equiv\Bigg\{\begin{array}{ll}
\xi^{J_l+s}_{\kappa_l-1} & \mbox{ if }\kappa_l\mbox{ is odd,}\\
i\xi^{J_l+s}_{\kappa_l-1} & \mbox{ if }\kappa_l\mbox{ is even.}\\
\end{array}
\end{equation}
Then due to Proposition \ref{Kprop2}, $A$ is symmetric, and so also is $A_2$, and hence $A_2$ must be diagonalizable. Then the following Lemma is true.
\begin{lem}\label{Klem2}
	\begin{equation}
	A_2(X^l_s)=\kappa_l(\kappa_l+1)X^l_s\mbox{   and   }A_2(Y^l_s)=0\mbox{ for }l=1,...,I\mbox{ and }s=0,1,...,m_l-1.
	\end{equation}
\end{lem}
In other words, the set $\{X^l_s,Y^l_s\,|\,l=1,...,I;\,s=0,1,...,m_l-1\}$ forms an eigenbasis of $A_2$. An immediate consequence of this is that $\mbox{spec}(A_2)=\{0\}\cup\{\kappa_j(\kappa_j+1)\,|\,j=1,...,I\}$, and $m_j$ is the dimension of the eigenspace associated to the eigenvalue $\kappa_j(\kappa_j+1)$ ($I$ being the number of distinct positive eigenvalues of $A_2$).

We now define the spaces
\begin{equation}\label{E_0q}
E^l_0\equiv\mbox{span}_\R\{Y^l_s\,|\,s=0,1,...,m_l-1\},\quad E^l_+\equiv\mbox{span}_\R\{X^l_s\,|\,s=0,1,...,m_l-1\},
\end{equation}
and
\begin{equation}\label{E_0}
E_0\equiv\bigoplus_{l=1}^{I}E^l_0,\qquad E_+\equiv\bigoplus_{l=1}^{I}E^l_+.
\end{equation}
Then $E_0=\mbox{ker}(A_2)$ and $E^l_+$ is the eigenspace of $A_2$ corresponding to the eigenvalue $\kappa_j(\kappa_j+1)$. Also, from Proposition \ref{Kprop1} (iv) we see that $V_2=E_0\oplus E_+$.

\begin{lem}\label{Klem3}
	Suppose $X\in V_2$. Then $X\in\bigoplus_{q=1}^lE^q_0\oplus E_+^q$ if and only if $\Omega^{\kappa_l}_+\ccdot X=0$.
\end{lem}

\begin{lem}\label{Klem4}
	Suppose $X\in V_2$. Then $X\in\bigoplus_{q=1}^lE^q_0\oplus E^q_+$ if and only if $\Omega_+^{\kappa_l+2}\ccdot c(X)=0$.
\end{lem}

\begin{lem}\label{Klem5}
	Let $\,\widetilde{\,\,}\,:\,\Z_{\geq -1}\rar\{1,2,...,I\}$ be the map defined by
	\begin{equation}
	\widetilde{-1}=\tilde{0}=1\quad\mbox{ and }\quad\tilde{s}=\max\,\{l\,|\,\kappa_l\leq s\}\mbox{ if }s>0.
	\end{equation}
	Then
	\begin{enumerate}
		\item $\kappa_{\tilde{s}}\leq s$ for every $s\in\Z_{\geq0}$,\\
		\item $\kappa_{\tilde{s}}\leq s\leq\kappa_{\tilde{s}+1}$ for every $s\in\{0,1,...,\kappa_{I-1}\}$.
	\end{enumerate}
\end{lem}

\begin{lem}\label{Klem6}
	If $X\in V_2$, $\kappa_{\tilde{p}}+s<\kappa_{\tilde{p}+1}\,(s\geq 0),$ and $\Omega_+^{\kappa_{\tilde{p}}+s}\ccdot X=0$, then $\Omega^{\kappa_{\tilde{p}}}_{+}\ccdot X=0$.
\end{lem}

The next theorem is the most important result in this Section: it is vital to the proof of local existence at the origin.

\begin{thr}\label{Kthr9}
	Suppose $p\in\{1,2,...,\kappa_I-1\}$ and $Z_0,Z_1,...,Z_{p+1}\in V_2$ is a sequence of vectors satisfying $Z_0\in E^1_0\oplus E^1_+$ and $Z_{n+1}\in\bigoplus_{q=1}^{\tilde{n}}E^q_0\oplus E^q_+$ for $n=0,1,...,p$. Then for every $j\in\{1,2,...,p+1\}$, $s\in\{0,1,...,j\}$,
	\begin{enumerate}
		\item $[[c(Z_{j-s}),Z_s],Z_{p+2-j}]\in\bigoplus_{q=1}^{\tilde{p}}E^q_0\oplus E^q_+$, \\
		\item $[[c(Z_{p+2-j}),Z_{j-s}],Z_s]\in\bigoplus_{q=1}^{\tilde{p}}E^q_0\oplus E^q_+$.
	\end{enumerate}
\end{thr}
\begin{prop}\label{Kprop4}
	Let $W_0$ be regular. Then if $\Omega_+\in\slim_{\alpha\in \Sigma_\lambda}\R\bv{\emph{e}}_{\alpha}$, $E_+=\slim_{\alpha\in \Sigma_\lambda}\R\bv{\emph{e}}_{\alpha}$.
\end{prop}

\subsubsection{Proof of local existence at the origin ($r=0$)}\label{sec_lex0}

Now we use Theorem \ref{Kthm3} and the results of Section \ref{sec:sl2c} to prove the existence of solutions, unique and analytic with respect to their boundary parameters, in some neighbourhood of the origin. We begin by introducing some necessary notation, which will be used throughout this Section. First, we define the set
\begin{equation}
\mc{E}\equiv\{\kappa_j\,|\,j=1,...,I\},
\end{equation}
for $\kappa_j$ given in \eqref{kapdef}; and a set of projection operators
\begin{equation}\label{pplus}
\mbox{p}_+^q\,:\,E_+\rar E_+^q\,\, (q=1,...,I),
\end{equation}
between the spaces defined in \eqref{E_0q} and \eqref{E_0}. Also, we define $\bv{I}_\epsilon(0)$ as an open interval of size $|2\epsilon|$ on the real line about the point $0\in\R$:
\begin{equation}
\bv{I}_\epsilon(0)\equiv(-\epsilon,\epsilon)
\end{equation}
where for our purposes, $\epsilon>0$ is small. 

Using Proposition \ref{Kprop4} and equation \eqref{YM20}, we know that the solution $W_+(r)$ of equation \eqref{YM} is completely characterised by the condition
\begin{equation}
W_+(r)\in E_+\,\,\,\forall r.
\end{equation}
We noted previously that equation \eqref{EE2} decouples from the others, so that once we have solved equations \eqref{EE1} and \eqref{YM} for $\mu$ and $W_+$, we may easily solve \eqref{EE2} to give $S$. However, for completeness, we shall include $S$ in our analysis. 

We now have everything we need to state our Proposition:
\begin{prop}\label{prop:lex0}
	In a neighbourhood of the origin $r=0$ (i.e. for solitons only), there exist regular solutions to the field equations, analytic and unique with respect to their initial values, of the form
	\begin{equation}\label{prop0exp}
	\begin{split}
	m(r)&=m_3r^3+O(r^4),\\
	S(r)&=S_0+O(r^2),\\
	\omega_i(r)&=\omega_{i,0}+\slim_{j=1}^\mc{L} Q_{ij}\hat{u}_j(r)r^{k_j+1},\quad i=1,...,\mc{L}.
	\end{split}
	\end{equation}
	Above, $Q_{ij}$ is a non-singular matrix for which the $j$th column is the eigenvector of the matrix $\emph{\bv{A}}$ \eqref{Adef} with eigenvalue $k_j(k_j+1)$, and $\hat{u}_j(r)$ are some functions of $r$. Each solution is entirely and uniquely determined by the initial values $\hat{u}_j(0)\equiv\beta_j$, for arbitrary values of $\beta_j$. Once these are determined, the metric functions $m(r)$ and $S(r)$ are entirely determined.
\end{prop}

\textbf{Proof} Since $W_+(r)\in E_+$, we introduce new functions $u_{k}(r)$ with
\begin{equation}\label{K1}
W_+(r)=\Omega_++\sum_{s\in\mc{E}}u_{s+1}(r)r^{s+1},
\end{equation}
with $\Omega_+=W_+(0)$ and $u_{s+1}(r)\in E_+^{\tilde{s}}\,\forall r,\,\forall s\in\mc{E}$. This transformation is clearly invertible since $E_+=\bigoplus_{q=1}^IE_+^q$. Define
\begin{equation}
\chi_{s+1}=\Bigg\{\begin{array}{ll}
1 & \mbox{ if }s\in\mc{E},\\
0 & \mbox{ otherwise.}\\
\end{array}
\end{equation}
Then we may write \eqref{K1} as $W_+(r)=\Omega_++\slim_{k=0}^{\infty}\chi_ku_k(r)r^k$. Substituting this into the YM equations \eqref{YM}, we find:
\begin{equation}\label{BigF}
\mc{F}=-\slim_{k\in\mc{E}}A_2(u_{k+1})r^{k+1}+\slim_{k=2}^{N_1}f_kr^k
\end{equation}
for some $N_1\in\mathbb{Z}$, and
\begin{equation}\label{fdef}
\begin{split}
f_k=&\frac{1}{2}\slim_{j=2}^{k-2}\bigg\{\left[\left[\Omega_+,c(\chi_ju_j)\right]+\left[\Omega_-,\chi_ju_j\right],\chi_{k-j}u_{k-j}\right]\\
&+\left[\left[\chi_ju_j,c(\chi_{k-j}u_{k-j})\right],\Omega_+\right]+\slim_{s=2}^{j-2}\left[\left[\chi_su_s,c(\chi_{j-s}u_{j-s})\right],\chi_{k-j}u_{k-j}\right]\bigg\}.
\end{split}
\end{equation}
The need for the results of Section \ref{sec:sl2c} becomes apparent if we examine those results alongside the forms of \eqref{BigF} and \eqref{fdef}. Now since $A_2(u_{k+1})=k(k+1)u_{k+1}$, \eqref{BigF} becomes
\begin{equation}\label{BigF2}
\mc{F}=-\slim_{k\in\mc{E}}k(k+1)u_{k+1}r^{k+1}+\slim_{k=2}^{N_1}f_kr^k.
\end{equation} 

We proceed by defining new variables $v_{s+1}\equiv u'_{s+1},\,\,\forall s\in\mc{E}$. The YM equations \eqref{YM} become
\begin{equation}\label{slimkinE}
\begin{split}
r\slim_{k\in\mc{E}}v'_{k+1}r^{k+1}=&-2\slim_{k\in\mc{E}}(k+1)v_{k+1}r^{k+1}+\slim_{k\in\mc{E}}\frac{k(k+1)}{r}\left(\frac{1}{\mu}-1\right)u_{k+1}r^{k+1}\\
&-\frac{2}{r\mu}\left(m-\frac{P}{r}+\frac{r^3}{\ell^2}\right)\slim_{k\in\mc{E}}\left(v_{k+1}r^{k+1}+(k+1)u_{k+1}r^{k+1}\right)\\
&-\frac{1}{\mu}\slim_{k=4}^{N_1}f_kr^{k-1}.
\end{split}
\end{equation}
Now we apply projection operators $\mbox{p}_+^{\tilde{k}}$ \eqref{pplus} to equations \eqref{slimkinE} for each $k\in\mc{E}$, giving
\begin{equation}\label{vprime}
\begin{split}
rv'_{k+1}=&-2(k+1)v_{k+1}-\frac{2}{r\mu}\left(m-\frac{P}{r}+\frac{r^3}{\ell^2}\right)v_{k+1}+\frac{k(k+1)}{r}\left(\frac{1}{\mu}-1\right)u_{k+1}\\
&-\frac{2}{r^2\mu}\left(m-\frac{P}{r}+\frac{r^3}{\ell^2}\right)(k+1)u_{k+1}-\frac{1}{r^{k+1}\mu}\slim_{s=2}^{N_1-2}\mbox{p}_+^{\tilde{k}}(f_{s+2})r^{s+1}\\
\end{split}
\end{equation}
for all $k\in\mc{E}$. The main obstacle to writing this equation in the correct form for Theorem \ref{Kthm3} is the final term, as was the case for $\sun$ \cite{baxter_existence_2008, kunzle_analysis_1994}. As written it contains terms of much lower order than we want, i.e. terms of order $r^{-s}$ where $s>0$. Happily we may rewrite the final term using the following equality:
\begin{equation}\label{frewrite}
\frac{1}{r^{k+1}\mu}\slim_{s=2}^{N_1-2}\mbox{p}_+^{\tilde{k}}(f_{s+2})r^{s+1}=\frac{1}{\mu}\slim_{s=k}^{N_1-2}\mbox{p}_+^{\tilde{k}}(f_{s+2})r^{s-k}.
\end{equation}
We make the derivation of this plain by using the results from Section \ref{sec:sl2c}. Using Proposition \ref{Kprop4} and equation \eqref{fdef}, we may show that $f_k\in E_+\,\,\forall k$. From how we have defined the functions $u_{s+1}(r)$, we may see that $\chi_{s+1}u_{s+1}\in\bigoplus_{q=1}^{\tilde{s}}E_+^q$ for $0\leq s\leq \kappa_I$. So let us use Theorem \ref{Kthr9}, taking $Z_0=\Omega_+$ and $Z_{k+1}=\chi_{k+1}u_{k+1}$ for $k\geq 0$. Then it is clear that $f_{s+2}\in\bigoplus_{q=1}^{\tilde{s}}E_+^q$. Hence,
\begin{equation}
\mbox{p}_+^{\tilde{k}}(f_{s+2})=0\mbox{   if }s<k,\,\,\forall k\in\mc{E},
\end{equation}
because if $k\in\mc{E}$, then $k=\kappa_{\tilde{k}}$ and so if $s<k=\kappa_{\tilde{k}}$, then $\tilde{s}<\tilde{k}$, proving \eqref{frewrite}.

Using \eqref{frewrite} in \eqref{vprime} and rearranging gives
\begin{equation}\label{rvprime}
\begin{split}
rv'_{k+1}=&-2(k+1)v_{k+1}-\frac{2}{r\mu}\left(m-\frac{P}{r}+\frac{r^3}{\ell^2}\right)v_{k+1}+\frac{k(k+1)}{r}\left(\frac{1}{\mu}-1\right)u_{k+1}\\
&-\frac{2}{r^2\mu}\left(m-\frac{P}{r}+\frac{r^3}{\ell^2}\right)(k+1)u_{k+1}-\frac{r}{\mu}\slim_{s=k}^{N_1-1}\mbox{p}_+^{\tilde{k}}(f_{s+3})r^{s-k}\\
&+\left(1-\frac{1}{\mu}\right)\mbox{p}_+^{\tilde{k}}(f_{k+2})-\mbox{p}_+^{\tilde{k}}(f_{k+2}),\quad\forall k\in\mc{E}.
\end{split}
\end{equation}

It is helpful to note that in this regime, $\frac{1}{\mu}-1=O(r^2)$. Using the properties of $\HHIP{\,}{\,}$ and the fact that $A_2(u_2)=2u_2$, we can show that there exist analytic functions
\begin{equation}\label{PG1}
\hat{P}:E_+\times\mathbb{R}\rar\mathbb{R},\quad\quad\hat{G}:E_+\times E_+\times\mathbb{R}\rar\mathbb{R},
\end{equation}
with
\begin{equation}
P=r^4\norm{u_2}^2+r^5\hat{P}(u,r),\quad\quad G=2r^2\norm{u_2}^2+r^3\hat{G}(u,v,r),
\end{equation}
and where $u=\sum_{s\in\mc{E}}u_{s+1}$, $v=\sum_{s\in\mc{E}}v_{s+1}$, and $\norm{X}^2=\lnorm X|X\rnorm$.

Now we rewrite the Einstein equations (\ref{EE1}, \ref{EE2}). We introduce a new mass variable
\begin{equation}\label{curlyMdef}
\mc{M}=\frac{1}{r^3}\left(m-r^3\norm{u_2}^2\right).
\end{equation}
(We know that $\norm{u_2}$ is always defined since $\kappa_1=1$ always and hence $1\in\mc{E}$.) Then (\ref{EE1}, \ref{EE2}) become
\begin{equation}\label{rMprime}
\begin{split}
r\mc{M}^\prime=&-3\mc{M}+r\left[\hat{P}(u,r)+\hat{G}(u,v,r)-2\lnorm u_2|v_2\rnorm\right.\\
&\left.-2r\left(\mc{M}+\norm{u_2}^2-\frac{1}{2\ell^2}\right)\left(2\norm{u_2}^2+r\hat{G}(u,v,r)\right)\right],\\
rS^\prime=&\,\;r^2S\left(4\norm{u_2}^2+2r\hat{G}(u,v,r)\right).
\end{split}
\end{equation}
We make one last variable change:
\begin{equation}\label{vhatdef}
\hat{v}_{k+1}=v_{k+1}+\frac{1}{2(k+1)}\mbox{p}_+^{\tilde{k}}(f_{k+2}).
\end{equation}

We proceed by fixing a vector $X\in E_+$ and define $\hat{v}=\sum_{s\in\mc{E}}\hat{v}_{s+1}$. Then from (\ref{rvprime}, \ref{curlyMdef}, \ref{vhatdef}), we can show there exists a neighbourhood $\mc{N}_X$ of $X\in E_+$, some $\epsilon>0$, and a sequence of analytic maps
\begin{equation}
\mc{G}_k:\mc{N}_X\times E_+\times \bv{I}_\epsilon(0)\times \bv{I}_\epsilon(0)\rar E_0^{\tilde{k}}\quad\forall k\in\mc{E},
\end{equation}
such that
\begin{equation}\label{rhatvprime}
r\hat{v}'_{k+1}=-2(k+1)\hat{v}_{k+1}+r\mc{G}_k(u,\hat{v},\mc{M},r).
\end{equation}
Also, with (\ref{rMprime}, \ref{vhatdef}) and using $v_{s+1}=u'_{s+1}$, there exist analytic maps
\begin{equation}
\begin{split}
\mc{H}_k: & E_+\times E_+\rar E_+^{\tilde{k}}\,\,\,\forall k\in\mc{E},\\
\mc{J}: & E_+\times E_+\times \mathbb{R}\times\mathbb{R}\rar\mathbb{R},\\
\mc{K}: & E_+\times E_+\times \mathbb{R}\times\mathbb{R}\rar\mathbb{R},
\end{split}
\end{equation}
such that
\begin{equation}\label{ruprime_etc}
\begin{split}
& ru'_{k+1}=r\mc{H}_k(u,\hat{v}),\\
& r\mc{M}^\prime=-3\mc{M}+r\mc{J}(u,\hat{v},\mc{M},r),\\
& rS^\prime=r^2\mc{K}(u,\hat{v},S,r).\\
\end{split}
\end{equation}

Now equations (\ref{rhatvprime}, \ref{ruprime_etc}) are in a form appropriate to Theorem \ref{Kthm3}. For fixed $X\in E_+$ there exists a unique solution $\{u_{k+1}(r,Y),\hat{v}_{k+1}(r,Y),\mc{M}(r,Y),S(r,Y)\}$, analytic in a neighbourhood of $(r,Y)=(0,X)$, satisfying
\begin{equation}\label{final0BCs}
\begin{split}
u_{s+1}(r,Y)&=Y_s+O(r)\quad\forall s\in\mc{E},\\
\hat{v}_{s+1}(r,Y)&=O(r)\quad\forall s\in\mc{E},\\
\mc{M}(r,Y)&=O(r),\\
S(r,Y)&=S_0+O(r^2),\\
\end{split}
\end{equation}
where $Y_s=\mbox{p}_+^{\tilde{s}}(Y)$. From the definition of $\mc{M}$ \eqref{curlyMdef}, we can show that $m(r)=O(r^3)$. Also, it is easy to see from (\ref{PG1}, \ref{vhatdef}, \ref{final0BCs}) that
\begin{equation}
P=O(r^4),\quad\quad\quad G=O(r^2).
\end{equation}
From the results of Section \ref{sec:sl2c}, there must exist an orthonormal basis $\{\bv{w}_j|j=1,...,\bsym{\Sigma}\}$ for $E_+$ consisting of the eigenvectors of $A_2$, i.e. $A_2(\bv{w}_j)=k_j(k_j+1)\bv{w}_j$. So we introduce new variables in this basis:
\begin{equation}\label{usexp}
\slim_{s\in\mc{E}}u_{s+1}(r)r^{s+1}=\slim_{j=1}^{\bsym{\Sigma}}\hat{u}_{j}(r)r^{k_j+1}\bv{w}_j.
\end{equation}
From Proposition \ref{Kprop1}, we know that $\bsym{\Sigma}=|\Sigma_\lambda|$, so we can write $\Sigma_\lambda=\{\alpha_j|j=1,...,\bsym{\Sigma}\}$; and from Proposition \ref{Kprop4}, we find that $\{\bv{e}_{\alpha_j}|j=1,...,\bsym{\Sigma}\}$ is also a basis for $E_+$. Therefore we can write
\begin{equation}\label{wvecdef}
\bv{w}_j=\slim_{k=1}^{\bsym{\Sigma}} Q_{kj}\bv{e}_{\alpha_k}.
\end{equation}
With this definition of the matrix $Q_{ij}$, it is clear that the columns of $Q_{ij}$ are the eigenvectors of $A_2$. Now we expand $\Omega_+$ and $W_+(r)$ in the same basis:
\begin{equation}\label{omegabase}
\Omega_+=\slim_{j=1}^{\bsym{\Sigma}}\omega_{j,0}\bv{e}_{\alpha_j},\quad\quad W_+(r)=\slim_{j=1}^{\bsym{\Sigma}}\omega_j(r)\bv{e}_{\alpha_j}.
\end{equation}
Then equations (\ref{K1}, \ref{usexp}, \ref{wvecdef}, \ref{omegabase}) imply that
\begin{equation}
\omega_i(r)=\omega_{i,0}+\slim_{j=1}^{\bsym{\Sigma}} Q_{ij}\hat{u}_j(r)r^{k_j+1},\quad i=1,...,\bsym{\Sigma},
\end{equation}
with $\omega^2_{i,0}=\lambda_i$. Finally, from \eqref{final0BCs} and \eqref{usexp} we obtain
\begin{equation}
\hat{u}_j(r,Y)=\beta_j(Y)+O(r),\quad j=1,...,\bsym{\Sigma},
\end{equation}
with $\beta_j(Y)\equiv\lnorm\bv{w}_j|Y\rnorm$. Therefore, we obtain the expansions \eqref{prop0exp}. $\Box$

\subsection{Proof of local existence at the event horizon $r=r_h$}\label{sec:lexrh}

Here, the situation is again quite similar to the asymptotically flat case \cite{oliynyk_local_2002}. Therefore, as was the case in \cite{oliynyk_local_2002}, we have no need of the results in Section \ref{sec:sl2c}. In particular, the space $E_+$ that we will use does not have to be of the form defined in \eqref{E_0} -- we may replace $E_+$ everywhere in the following with $\sum_{\alpha\in\Sigma_\lambda}\R\bv{e}_\alpha$, and it is not necessary to know that $E_+=\sum_{\alpha\in\Sigma_\lambda}\R\bv{e}_\alpha$ (which is the essence of Proposition \ref{Kprop4}). Thus, we use the notation $E_+$ purely for convenience.

We begin by introducing the variable
\labeq{\rho=r-r_h,}{rhodef}
so that for $r\rar r_h$ we are considering the limit $\rho\rar0$. Keeping in mind the boundary conditions in Section \ref{ssec:bcsrh}, we prove the following Proposition:

\begin{prop}\label{prop:lexrh}
	In a neighbourhood of the event horizon $r=r_h\neq0$ (i.e. $\rho=0$), there exist regular black hole solutions to the field equations \eqref{EE1} -- \eqref{YM}, analytic and unique with respect to their initial values, of the form
	\begin{equation}\label{proprhexp}
	\begin{split}
	\mu(\rho)&=\mu^\prime_h\rho+O(\rho^2),\\
	S(\rho)&=S_h+O(\rho),\\
	\omega_j(\rho)&=\omega_{j,h}+O(\rho),\\
	\end{split}
	\end{equation}
	where $\mu^\prime_h>0$.
\end{prop}

\textbf{Proof} Along with \eqref{rhodef}, we introduce some new variables:
\begin{subequations}
	\begin{align}
	\mu&=\rho(\bar{\lambda}+\nu),\label{newmu}\\
	V_+&=(\bar{\lambda}+\nu)W^\prime_+,\label{Vplus}
	\end{align}
\end{subequations}
for $\bar{\lambda}$, $V_+$ functions of $\rho$, and $\nu$ some constant yet to be determined. Immediately we have
\begin{equation}\label{rhodw}
\rho\frac{dW_+}{d\rho}=\rho\left(\frac{V_+}{\bar{\lambda}+\nu}\right),
\end{equation}
and it is clear that there exist analytic maps $\hat{\mc{F}}:E_+\rar E_+$, $\hat{P}:E_+\rar\mathbb{R}$, with
\begin{equation}
\hat{\mc{F}}(W_+)=\mc{F},\quad\quad\hat{P}(W_+)=P.
\end{equation}
Define an analytic map $\hat{G}:E_+\times \bv{I}_{|\nu|}(0)\rar\mathbb{R}$ by
\begin{equation}
\hat{G}(X,a)=\frac{1}{2(a+\nu)^2}\|X\|^2.
\end{equation}
Then we can see that $G=\hat{G}(V_+,\bar{\lambda})$. Using these we can rewrite the EYM equations \eqref{EE1} to \eqref{YM} as
\begin{subequations}\label{dlamdv}
	\begin{align}
	\rho\frac{d\bar{\lambda}}{d\rho}=&-(\bar{\lambda}+\nu)+\frac{1}{r_h}-\frac{2}{r_h^3}\hat{P}(W_+)+\frac{3r_h}{\ell^2}+\rho\left[\frac{3}{\ell^2}+\frac{1}{\rho}\left(\frac{1}{\rho+r_h}-\frac{1}{r_h}\right)\right.\nonumber\\
	&\left.-\frac{2}{\rho}\left(\frac{1}{(\rho+r_h)^3}-\frac{1}{r_h^3}\right)\hat{P}(W_+)+\left(\frac{\bar{\lambda}+\nu}{\rho+r_h}\right)\left(1+2\hat{G}(V_+,\bar{\lambda})\right)\right],\\
	\rho\frac{dV_+}{d\rho}=&-V_+-\frac{1}{(\rho+r_h)^3}\hat{\mc{F}}(W_+)-\rho V_+\left(\frac{2\hat{G}(V_+,\bar{\lambda})}{\rho+r_h}\right),\\
	\rho\frac{dS}{d\rho}=&\,\rho\,\frac{2S\hat{G}(V_+,\bar{\lambda})}{\rho+r_h}.
	\end{align}
\end{subequations}

In order to cast the equations in the form necessary for Theorem \ref{Kthm3}, we introduce some final new variables:
\begin{subequations}\label{newvarrh}
	\begin{align}
	\hat{\lambda}=&\bar{\lambda}+\nu-\frac{1}{r_h}+\frac{2}{r_h^3}\hat{P}(W_+)-\frac{3r_h}{\ell^2},\label{lamhat}\\
	\hat{V_+}=& V_++\frac{1}{r_h^3}\hat{\mc{F}}(W_+).
	\end{align}
\end{subequations}
We continue by defining an analytic map $\gamma:E_+\times\mathbb{R}\rar\mathbb{R}$ with
\begin{equation}
\gamma(X,a)=a-\nu+\frac{1}{r_h}-\frac{2}{r_h^3}\hat{P}(X)+\frac{3r_h}{\ell^2}.
\end{equation}
Fix a vector $Z\in E_+$ satisfying $\|r_h^{-1}-2r_h^{-3}\hat{P}(Z)+3r_h\ell^{-2}\|>0$. Then if we set
\begin{equation}
\nu=\frac{1}{r_h}+\frac{3r_h}{\ell^2}-\frac{2}{r_h^3}\hat{P}(Z),
\end{equation}
it is obvious that $\gamma(Y,0)=0$. Therefore, define an open neighbourhood $D$ of $(Z,0)\in E_+\times\mathbb{R}$ by
\begin{equation}
D=\{(X,a)\,|\,\|\gamma(X,a)\|<\|\nu\|\}.
\end{equation}
Then from (\ref{rhodw}, \ref{dlamdv}, \ref{newvarrh}) we can show there must exist some $\epsilon>0$ and analytic maps
\begin{equation}
\begin{split}
\mc{G}:& E_+\times D\rar\mathbb{R},\\
\mc{H}:& E_+\times D\times \bv{I}_\epsilon(0)\rar\mathbb{R},\\
\mc{J}:& E_+\times D\times \bv{I}_\epsilon(0)\rar\mathbb{R},\\
\mc{K}:& E_+\times\R\times\bv{I}_\epsilon(0)\rar\R,\\
\end{split}
\end{equation}
such that
\begin{equation}\label{rhfin}
\begin{split}
&\rho\frac{dW_+}{d\rho}=\rho\mc{G}(\hat{V_+},W_+,\hat{\lambda}),\\
&\rho\frac{d\hat{V_+}}{d\rho}=-\hat{V_+}+\rho\mc{H}(\hat{V_+},W_+,\hat{\lambda},\rho),\\
&\rho\frac{d\hat{\lambda}}{d\rho}=-\hat{\lambda}+\rho\mc{J}(\hat{V_+},W_+,\hat{\lambda},\rho),\\
&\rho\frac{dS}{d\rho}=\rho\mc{K}(\hat{V_+},S,\rho).\\
\end{split}
\end{equation}
It can be seen that equations \eqref{rhfin} are in the form applicable to Theorem \ref{Kthm3}. Hence there is a unique solution $\{W_+(\rho,Y),\hat{V_+}(\rho,Y),\hat{\lambda}(\rho,Y), S(\rho,Y)\}$, analytic in a neighbourhood of $(\rho,Y)=(0,Z)$, which satisfies
\begin{subequations}
	\begin{align}
	W_+(\rho,Y)&=Z+O(\rho),\label{finw}\\
	\hat{V_+}(\rho,Y)&=O(\rho),\label{finVhat}\\
	\hat{\lambda}(\rho,Y)&=O(\rho),\label{finlamhat}\\
	S(\rho,Y)&=S_h+O(\rho).
	\end{align}
\end{subequations}
To gain a more explicit solution, we expand $Z$, $W_+$ in the basis $\{\bv{e}_{\alpha_j} | j=1,...,\bsym{\Sigma}\}$, as follows:
\begin{equation}\label{rhbase}
Z=\slim_{j=1}^{\bsym{\Sigma}}\omega_{j,h}\bv{e}_{\alpha_j},\quad\quad W_+=\slim_{j=1}^{\bsym{\Sigma}}\omega_{j}(\rho)\bv{e}_{\alpha_j}.
\end{equation}
Noting \eqref{finw}, this yields
\begin{equation}
\omega_j(\rho,Z)=\omega_{j,h}+O(\rho)\quad\forall j=1,...,\bsym{\Sigma}.
\end{equation}

Finally, it is easy to show from (\ref{newmu}, \ref{lamhat}, \ref{finlamhat}) that
\begin{equation}
\mu(\rho,Z)=\nu\rho+O(\rho^2),
\end{equation}
and hence
\labeq{\mu_h=0,\qquad \mu'_h=\nu.}{}
Therefore, we have obtained the expansions \eqref{proprhexp}. $\Box$

\subsection{Proof of local existence as $r\rar\infty$}\label{sec_lexinf}

The behaviour of solutions in the asymptotic limit is the biggest difference between the asymptotically flat and adS cases. Because of the constraints on the asymptotic values of the gauge functions for $\Lambda=0$, the proof followed a similar route to the local existence at the origin. However for $\Lambda<0$, our situation is much more similar to the local existence at the event horizon, so we follow a similar method to that used in Proposition \ref{prop:lexrh} from Section \ref{sec:lexrh}. Hence, the same comments apply as at the beginning of Section \ref{sec:lexrh}: we do not need any of the results of Section \ref{sec:sl2c} here, and thus we use the notation $E_+$ out of utility.

To deal sensibly with the limit $r\rar\infty$ we transform to the variable 
\begin{equation}\label{zr}
z=r^{-1},
\end{equation}
whence we are now dealing with the limit $z\rar0$. We state our Proposition:

\begin{prop}\label{prop:lexinf}
	There exist regular solutions of the field equations in some neighbourhood of $z=0$, analytic and unique with respect to their initial values, of the form
	\begin{equation}\label{propinfexp}
	\begin{split}
	m(z)&=M+O(z),\\
	S(z)&=1+O(z^4),\\
	\omega_j(z)&=\omega_{j,\infty}+c_jz+O(z^2),\\
	\end{split}
	\end{equation}
	for arbitrary constants $\omega_{j,\infty}$, $c_j$; where in order to agree with the asymptotic limit of adS space, we have let $m_\infty=M$, the ADM mass of the solution, and $S_\infty=1$.
\end{prop}

\textbf{Proof} As well as \eqref{zr}, we introduce also the following new variables:
\begin{subequations}
	\begin{align}
	\lambda(z)&\equiv 2m(r),\\
	v_+(z)&\equiv r^2W^\prime_+(r).\label{infvars}
	\end{align}
\end{subequations}
We immediately find that
\begin{equation}
z\frac{dW_+}{dz}=-zv_+,
\end{equation}
and it is clear that there exist analytic maps $\hat{\mc{F}}:E_+\rar E_+$ and $\hat{P}:E_+\rar\mathbb{R}$ with
\begin{equation}
\hat{\mc{F}}(W_+)=\mc{F},\quad\quad\hat{P}(W_+)=P.
\end{equation}
Also we find that
\begin{equation}
G=\frac{z^4}{2}(v_+,v_-),
\end{equation}
which means that
\begin{equation}
z\frac{dS}{dz}=-z^4\norm{v_+}^2S.
\end{equation}

For $\lambda$ and $v_+$, it can be shown that
\begin{equation}\label{infeqs}
\begin{split}
z\frac{d\lambda}{dz}&=-z\left(2\hat{P}(W_+)+\norm{v_+}^2\left(z^2-\lambda z^3+\frac{1}{\ell^2}\right)\right),\\
z\frac{dv_+}{dz}&=2v_+\left(\frac{1}{\mu z^2\ell^2}-1\right)+\frac{1}{\mu z}\left(\hat{\mc{F}}(W_+)+z^2v_+\left(\lambda-2\hat{P}(W_+)z\right)\right).\\
\end{split}
\end{equation}
It is useful to note that in the asymptotic limit, $\mu\sim 1+\frac{1}{z^2\ell^2}$, from which we may see that
\begin{equation}\label{mulim}
\frac{1}{\mu z^2\ell^2}-1=O(z^2),\quad\mbox{ and }\quad \frac{1}{\mu z}=O(z).
\end{equation} 

Examining the number of degrees of freedom we expect at this boundary, we fix two vectors $X,C\in E_+$. Then from results \eqref{infvars} -- \eqref{mulim}, it is clear that there exists an $\epsilon>0$ and analytic maps
\begin{equation}
\begin{split}
\mc{G}_\infty:& E_+\rar\R,\\
\mc{H}_\infty:& \,E_+\times\R\rar\R,\\
\mc{J}_\infty:& \,E_+\times E_+\times\R\times\bv{I}_\epsilon(0)\rar\R,\\
\mc{K}_\infty:& \,E_+\times E_+\times\R\times\bv{I}_\epsilon(0)\rar\R,\\
\end{split}
\end{equation}
with
\begin{subequations}\label{inffin}
	\begin{align}
	& z\frac{dW_+}{dz}=z\mc{G}_\infty(v_+),\\
	& z\frac{dS}{dz}=z^4\mc{H}_\infty(v_+,S),\\
	& z\frac{d\lambda}{dz}=z\mc{J}_\infty(W_+,v_+,\lambda,z),\\
	& z\frac{dv_+}{dz}=z\mc{K}_\infty(W_+,v_+,\lambda,z)\label{dv+}
	\end{align}
\end{subequations}

(noting that $\mc{G}_\infty$ is just the map $v_+\mapsto-v_+$). Now we are at the stage where we may apply Theorem \ref{Kthm3}; and hence it is clear that these equations possess a unique solution $\{S(z,Y,Z),\lambda(z,Y,Z),W_+(z,Y,Z),v_+(z,Y,Z)\}$ analytic in some neighbourhood of $(z,Y,Z)=(0,X,C)$ with behaviour
\begin{subequations}
	\begin{align}
	S(z,Y,Z)&=S_\infty+O(z^4),\\
	\lambda(z,Y,Z)&=\lambda_\infty+O(z),\\
	W_+(z,Y,Z)&=X+O(z),\label{wexp}\\
	v_+(z,Y,Z)&=C+O(z).\label{vplusc}
	\end{align}
\end{subequations}

However, noting \eqref{zr} and \eqref{infvars}, we may integrate \eqref{vplusc}, choosing the constant (vector) of integration to agree with \eqref{wexp}. This combines \eqref{wexp} and \eqref{vplusc}, yielding
\begin{equation}
W_+(z,Y,Z)=X-Cz+O(z^2).
\end{equation}
To gain an explicit solution in terms of the components of $X$, $C$ and $W_+$, we expand them all in the same basis:
\begin{equation}\label{XWbases}
W_+=\slim_{\alpha\in\Sigma_\lambda}\omega_{\alpha}(z)\bv{e}_{\alpha}, \quad X=\slim_{\alpha\in\Sigma_\lambda}\omega_{\alpha,\infty}\bv{e}_{\alpha},\quad C=\slim_{\alpha\in\Sigma_\lambda}(-c_\alpha)\bv{e}_\alpha.
\end{equation}
No constraints are placed on the constants $\omega_{\alpha,\infty}$ or $c_\alpha$. Then it is clear that near $z=0$, the gauge field functions have the form
\begin{equation}\label{omal}
\omega_{\alpha}(z)=\omega_{\alpha,\infty}+c_\alpha z+O(z^2),\quad\forall \alpha\in\Sigma_\lambda.
\end{equation}
Finally, noting that we expect our solution to approach adS space in the asymptotic limit, we set $\lambda_\infty\equiv 2M$, $S_\infty\equiv1$, and thus recover the expansions \eqref{propinfexp}. $\Box$

\section{Global existence arguments}\label{sec:gloex}

Now we turn our attention to proving the existence of global solutions to our field equations. Here we have a choice of approaches. We considered using the more novel approach of Nolan and Winstanley \cite{nolan_existence_2012} who let the initial conditions and embedded solutions reside in appropriate Banach spaces, and then recast the field equations so that they could apply the Implicit Function Theorem, hence proving that non-trivial solutions exist in some neighbourhood of embedded solutions. However, it appears to be necessary to their argument that $m(r)$ is constant for the embedded solution, something we have not been able to get around yet, meaning that we could only identify solutions in a neighbourhood of the embedded SadS solution.

Alternatively, the traditional argument that has been used in this case is the `shooting argument' (used in e.g. \cite{baxter_existence_2015, winstanley_existence_1999}), which basically involves proving the existence of solutions locally at the boundaries, and then proving that solutions which begin at the initial boundary $r=r_h$ ($r=0$) near to existing embedded solutions can be integrated out arbitrarily far, remaining regular right into the asymptotic regime, where they will `meet up' with solutions existing locally at $r\rar\infty$; and that these neighbouring solutions will remain close to the embedded solution. While this seems somehow less elegant, there are no restrictions on the embedded solution we may use, and hence the proof we are able to create is more general and hence more powerful. Therefore, we resign ourselves to using the more traditional techniques.

We begin by noting that we have already considered the behaviour of the field equations in the asymptotic limit and shown that solutions will in general remain regular in this regime (Section \ref{sec:asym}), so we must now make sure that any solution which begins regularly at the initial boundary $r=r_h$ ($r=0$) can be integrated out arbitrarily far while the field variables remain regular. 

\begin{prop}\label{prop:gloreg}
	If $\mu(r)>0\,\,\,\forall r\in[r_h,\infty)$ for black holes, or $\forall r\in[0,\infty)$ for solitons, then all field variables may be integrated out from the boundary conditions at the event horizon (or the origin) into the asymptotic regime, and will remain regular.
\end{prop}

\textbf{Proof} Define $\mc{Q}\equiv[r_0,r_1)$ and $\bar{\mc{Q}}\equiv[r_0,r_1]$, where $r_0=r_h$ for black holes and $r_0=0$ for solitons, and $r_0<r_1<\infty$. Our strategy is to assume that all field variables are regular on $\mc{Q}$, i.e. in a neighbourhood of $r=r_0$, and then show using the field equations that as long as the metric function $\mu(r)>0\,\,\forall r\in[r_0,\infty)$, then they will remain regular on $\bar{\mc{Q}}$ also, i.e. at $r=r_1$; and thus we can integrate the field equations out arbitrarily far and the field variables will remain regular.

First notice that $G,P>0$ by the definitions \eqref{quantdefs}. This means using \eqref{EE1} that $m'(r)>0\,\,\forall r$ and thus $m(r)$ is monotonic increasing, as expected for the physical mass. This means that (if it exists),
\begin{equation}\label{mmax}
m_{\max}\equiv\sup\{m(r)\,|\,r\in\bar{\mc{Q}}\}=m(r_1).
\end{equation}
The same applies to $(\ln |S(r)|)^\prime$ (see \eqref{EE2}), showing that $\ln|S(r)|$ and hence $S(r)$ is monotonic increasing too, so that (again, if we can prove that $S$ is finite on $\bar{\mc{Q}}$)
\begin{equation}
S_{\max}\equiv\sup\{S(r)\,|\,r\in\bar{\mc{Q}}\}=S(r_1).
\end{equation}

The condition $\mu(r)>0\quad\forall r\in[r_0,\infty)$ gives us our starting point, since this implies that
\begin{equation}
m(r_1)\leq \frac{r_1}{2}+\frac{r_1^3}{2\ell^2},
\end{equation}
giving us an absolute upper bound to work with. This in turn implies that $m(r)$ is bounded on $\bar{\mc{Q}}$ (and so \eqref{mmax} holds), and thus also that $\mu(r)$ is bounded on $\bar{\mc{Q}}$. Thus we may define $\mu_{\min}\equiv\inf\{\mu(r)\,|\,r\in\bar{\mc{Q}}\}$.

Now we examine \eqref{EE1}. It is clear that
\begin{equation}
2m^\prime(r)\geq 2\mu G,
\end{equation}
and integrating, we can show that
\begin{equation}\label{Gbound}
\frac{2[m(r_1)-m(r_0)]}{\mu_{\min}}\geq 2\int\limits_{r_0}^{r_1}Gdr,
\end{equation}
which implies from \eqref{EE2} that $\ln|S|$ and hence $S$ is bounded on $\bar{\mc{Q}}$. 

Equation \eqref{Gbound} also implies that $G$ is bounded on $\bar{\mc{Q}}$, and since
\begin{equation}
2G=\norm{W_+^\prime}^2,
\end{equation}
then again by integrating and using the Cauchy-Schwartz inequality,
\begin{equation}
\int\limits^{r_1}_{r_0}2Gdr=\int\limits^{r_1}_{r_0}\norm{W_+^\prime}^2dr\geq\left(\int\limits^{r_1}_{r_0}\norm{W_+}^\prime dr\right)^{\!\!2},
\end{equation}
and hence
\begin{equation}
\int\limits^{r_1}_{r_0}2Gdr\geq\left(\norm{W_+}\Big|_{r=r_1}-\norm{W_+}\Big|_{r=r_0}\right)^{\!\!2}.
\end{equation}
The left hand side is bounded, and the right hand side is a sum of positive terms and hence bounded below by 0. Thus $\norm{W_+}$ and hence $W_+$ is bounded on $\bar{\mc{Q}}$. Since $W_0$ is constant and $W_-=-c(W_+)$, this also means that $\hat{F}$ and hence $\mc{F}$ and $P$ are similarly bounded on $\bar{\mc{Q}}$ (see \eqref{quantdefs}).

Finally, we may rewrite the YM equations \eqref{YM} as
\begin{equation}
\left(\mu SW^\prime_+\right)^\prime=-\frac{S\mc{F}}{r^2}.
\end{equation}
Integrating and rearranging gives
\begin{equation}
\mu(r_1)S(r_1)W_+^\prime(r_1)=\mu(r_0) S(r_0)W_+^\prime(r_0)-\int\limits_{r_0}^{r_1}\frac{S\mc{F}}{r^2}dr,
\end{equation}
and since all functions on the right hand side are bounded on $\bar{Q}$ (see \eqref{quantdefs}), as are $\mu$ and $S$, then we can finally conclude that $W_+^\prime$ is bounded on $\bar{\mc{Q}}$. $\Box$

\subsection{Global existence of solutions in a neighbourhood of embedded solutions}

Finally, we may prove the major conclusions of our research, which hinge on the following Theorem. The gist of it is that global solutions to the field equations \eqref{EE12} -- \eqref{YM2}, which we have proven are uniquely characterised by the appropriate boundary values and analytic in those values, exist in open sets of the initial parameter space; and hence that solutions which begin sufficiently close to existing solutions to the field equations will remain close to them as they are integrated out arbitrarily far into the asymptotic regime, remaining regular throughout the range. It can be noted that this argument is quite similar to those we have used for the $\sun$ case \cite{baxter_existence_2008, baxter_existence_2016}.

\begin{thr}\label{prop:gloex}
	Assume we have an existing solution of the field equations \eqref{EE12} to \eqref{YM2}, with each gauge field function $\omega_j(r)$ possessing $n_j$ nodes each, and with initial gauge field values $\{\omega_{1,0},\omega_{2,0},...,\omega_{\mc{L},0}\}$, taking $\{\omega_{j,0}\}=\{\omega_{j,h}\}$ for black holes and $\{\omega_{j,0}\}=\{\beta_j\}$ for solitons. Then all initial gauge field values $\{\tilde{\omega}_{j,0}\}$ in a neighbourhood of these values will also give a solution to the field equations in which each gauge field function $\tilde{\omega}_j(r)$ has $n_j$ nodes.
\end{thr}
\vspace{5mm}
\textbf{Proof} Assume we possess an existing solution to the field equations \eqref{EE12} to \eqref{YM2}, where each gauge function $\omega_j(r)$ has $n_j$ nodes and initial conditions $\omega_{j,0}\neq 0$ in general. Proposition \ref{prop:gloreg} and the analysis in Section \ref{sec:asym} show that as long as $\mu(r)>0$ we may integrate this solution out arbitrarily far into the asymptotic regime to obtain a solution which will satisfy the boundary conditions as $r\rar\infty$. For the rest of the argument, we assume that $\ell$ is fixed and so is $r_h$ for black holes and that each gauge function $\omega_j$ has $n_j$ nodes.

From the local existence results (Propositions \ref{prop:lex0}, \ref{prop:lexrh} and \ref{prop:lexinf}), we know that for any set of initial values, solutions exist locally near the event horizon for a black hole, or the origin for a soliton, and that they are analytic in their choice of initial conditions. Again we use the notation $r_0=r_h$ for black holes and $r_0=0$ for solitons. For an existing solution, it must be true that $\mu(r)>0$ for all $r\in[r_0,\infty)$. So, by analyticity, all sufficiently nearby solutions will also have $\mu(r)>0$ for all $r\in[r_0,r_1]$ for some $r=r_1$ with $r_0<r_1<\infty$. By Proposition \ref{prop:gloreg}, this nearby solution will also be regular on $[r_0,r_1]$.

Now, let $r_1 >> r_0$, so that for the existing solution, $m(r_1)/r_1<<1$. Let $\{\tilde{\omega}_{j,0}\}$ be a different set of initial conditions at $r=r_0$ for gauge fields $\tilde{\omega}_j$, such that $\{\tilde{\omega}_{j,0}\}$ are in some small neighbourhood of $\{\omega_{j,0}\}$; and let $\tilde{m}(r)$ be the mass function and $\tilde{\mu}$ be the metric function of that solution. By analyticity (as above), $\tilde{\mu}(r)>0$ on this interval, so this new solution will also be regular on $[r_0,r_1]$; and since the two solutions must remain close together, the gauge functions $\tilde{\omega}_j$ will also each have $n_j$ nodes.

Also it is then the case that $\tilde{m}(r_1)/r_1<<1$, and since $r_1 >> r_0$ we consider this the asymptotic regime. Provided $r_1$ is large enough (and hence $\tau_1$ is very small), the solution will not move very far along its phase plane trajectory as $r_1\rar\infty$ (see Section \ref{sec:asym}). Therefore $\tilde{m}(r)/r$ remains small, the asymptotic regime remains valid, and the solution will remain regular for $r$ arbitrarily large. $\Box$

\begin{cor}
	Non-trivial solutions to the field equations which are nodeless, i.e. for which $\omega_j(r)\neq0\quad\forall r$, exist in some neighbourhood of both existing trivial SadS solutions (described in \ref{sec:SadS}), and embedded $\essu$ solutions (proven in Proposition \ref{prop:su2embed}).
\end{cor}

\subsection{Existence of solutions in the large $|\Lambda|$ limit ($\ell\rar 0$)}\label{sec:l0}

So far we have proven the existence of global black hole and soliton solutions in some neighbourhood of existing solutions, for fixed $r_h$ and $\Lambda$. But there is a further consideration, revealed by investigations into $\sun$. On the one hand, we discovered numerically that as $N$ increases, regions of the parameter space in which we may find nodeless solutions shrink in size \cite{baxter_abundant_2008, baxter_existence_2008}; on the other, for $|\Lambda|$ large enough, \textit{all} solutions we found were nodeless. In addition, when we investigated the linear stability of these solutions \cite{baxter_stability_2015}, we were only able to prove stability in the limit $|\Lambda|\rar\infty$, due to terms arising in the gravitational sector.

In view of the similarities between the case under consideration and the $\sun$ case, it is sensible to investigate this limit in the case of a general compact gauge group. Our strategy is to transform the field variables such that we may sensibly find a unique solution to the equations at $\ell=0$. Then, noting that it is only in the asymptotic limit that the influence of $\ell$ is felt, we modify Proposition \ref{prop:lexinf} using our new variables, and show that the arguments used in Section \ref{sec:gloex} may be easily adapted to serve in a neighbourhood of $\ell=0$. 

We must emphasise that we cannot prove the existence of global non-trivial solutions \emph{at} $\ell=0$, since in that case the asymptotic variable we used in Section \ref{sec:asym} becomes singular and therefore that part of the proof breaks down.

\begin{thr}\label{thr:l0}
	There exist non-trivial solutions to the field equations \eqref{EE1} -- \eqref{YM}, analytic in some neighbourhood of $\ell=0$, for any choice of boundary gauge field values. For black holes, these are given by $\{\omega_{j,h}\}$ ($j=1,...,\mc{L}$) (in the base \eqref{rhbase}); for solitons, $\{\beta_j\}$, ($j=1,...,\mc{L}$).
\end{thr}

\textbf{Proof} We'll take the black hole case to begin with, noting that we fix $r_h$ for the rest of the argument. Let us change to the variables
\begin{subequations}
	\begin{align}
	\bar{m}&=m\ell^2,\label{barm}\\
	W_{\pm}^\prime&=\ell\sqrt{2}X_{\pm}\label{Xpm}.
	\end{align}
\end{subequations}
The field equations \eqref{EE1} -- \eqref{YM} then become
\begin{equation}
\begin{split}
\frac{d\bar{m}}{dr}&=\ell^2\left[\left(\ell^2-\frac{2\bar{m}}{r}+r^2\right)\norm{X_+}^2-\frac{P}{2r^2}\right],\\
\frac{1}{S}\frac{dS}{dr}&=\frac{2\ell^2}{r}\norm{X_+^\prime}^2,\\
0&=r^2\left(\ell^2-\frac{2\bar{m}}{r}+r^2\right)X_+^\prime+\left(2\bar{m}-\frac{P\ell^2}{r}+2r^3\right)X_++\ell\mc{F}.\\
\end{split}
\end{equation}
Taking the (now allowed) limit $\ell\rar0$:
\begin{equation}
\begin{split}
\frac{d\bar{m}}{dr}&=0,\\
\frac{1}{S}\frac{dS}{dr}&=0,\\
0&=r^2\left(-\frac{2\bar{m}}{r}+r^2\right)X_+^\prime+\left(2\bar{m}+2r^3\right)X_+.\\
\end{split}
\end{equation}

The first of these is easily integrated to give $\bar{m}$ constant, which we therefore set to $\bar{m}(r)=\bar{m}_h$. We also notice that since
\begin{equation}
\bar{m}_h=\ell^2m_h=\frac{\ell^2r_h}{2}+\frac{r_h^3}{2},
\end{equation}
then we must have $\bar{m}(r)=\frac{r_h^3}{2}$ at $\ell=0$. The second integrates to $S$ constant, which we set to 1 in agreement with the asymptotic limit. The third is readily integrated to give
\begin{equation}
X_+(r)=\frac{\mc{X}r}{r^3-r_h^3},
\end{equation}
for $\mc{X}$ a constant of integration. However this is singular at both $r=r_h$ and as $r\rar\infty$ unless we take $\mc{X}=0$, giving $X_+(r)\equiv 0$. Examining \eqref{Xpm} and noting that we will want to vary this solution away from $\ell=0$ to small non-zero values of $\ell$, we see that $W_+(r)$ is also a constant, for which we are forced to take $W_+(r)\equiv W_+(r_h)$.

Hence using an appropriate basis for $W_+(r)$ \eqref{Wplus}, the unique solution obtained is
\begin{equation}\label{solnl0}
\bar{m}(r)\equiv\frac{r_h^3}{2},\qquad S(r)\equiv 1,\qquad \omega_\alpha(r)\equiv\omega_{\alpha,h},\,\,\,\forall\alpha\in\Sigma_\lambda.
\end{equation}

We note that this is identical to the $\sun$ case. 

Now we take Proposition \eqref{prop:lexinf} and re-purpose it to the case at hand. Defining new variables
\begin{equation}
\tilde{\lambda}\equiv\lambda\ell^2,\qquad\tilde{\mu}\equiv\mu\ell^2,
\end{equation}
the field equations \eqref{infeqs} become
\begin{equation}
\begin{split}
z\frac{d\tilde{\lambda}}{dz}&=-z\left(2\ell^2\hat{P}(W_+)+\norm{v_+}^2\left(\ell^2 z^2-\tilde{\lambda}z^3+1\right)\right),\\
z\frac{dv_+}{dz}&=2v_+\left(\frac{1}{\tilde{\mu} z^2}-1\right)+\frac{\ell^2}{\tilde{\mu} z}\left(\hat{\mc{F}}(W_+)+z^2v_+\left(\tilde{\lambda}-2\hat{P}(W_+)z\right)\right);\\
\end{split}
\end{equation}
and the equation for $S$ is unchanged. But the structure of the field equations is unaltered, and so the proof given in Section \ref{sec_lexinf} is unchanged. Then, for arbitrarily small $\ell$, we may find solutions that exist locally in the asymptotic limit.

The argument that proves that non-trivial global solutions exist for small $\ell$ is very similar to Proposition \ref{prop:gloex}. We fix $r_h$, take the existing solution \eqref{solnl0}, and consider varying $\{\omega_{j,h}\}$, and varying $\ell$ away from 0. Note that for the embedded solution \eqref{solnl0}, all gauge fields will be nodeless. We then choose some $r_1 >> r_h$ so that we can consider $r_1$ in the asymptotic regime. Proposition \ref{prop:lexrh} confirms that for $\ell$ sufficiently small we can find solutions near the existing unique solution which will begin regularly near $r = r_h$ and remain regular also at $r=r_1$, and that those solutions will have nodeless gauge field functions due to analyticity. Finally, since we are now in the asymptotic regime, we can use the logic in Section \ref{sec:asym} and Proposition \ref{prop:gloreg} to ensure that solutions will remain regular as $r\rar\infty$ and that all $\omega_j$ will be nodeless. 

The corresponding proof for solitons is similar to that for black holes, though we must be more careful about how we take the limit $\ell\rar0$. The parameter $\tau\propto r^{-1}$ that we use in the asymptotic regime is fine for black holes since $\min\{r\}=r_h$ so $\tau$ is bounded and thus $r^{-1}$ remains regular throughout the range $[r_h,\infty)$; but this is clearly no longer the case for solitons as $\min\{r\}=0$ so that $\tau$ becomes singular.

We follow the clues in the $\sun$ case \cite{baxter_existence_2008} and rescale all dimensionful quantities:
\begin{equation}\label{l0var}
r=\ell x,\qquad\qquad m(r)=\ell\check{m}(x).
\end{equation}
In addition, we find it best to work with the gauge functions $\hat{u}_j(r)$ which we defined in the proof of local existence at the origin, Proposition \ref{prop:lex0}, using
\begin{equation}\label{omega0}
\omega_i(x)=\omega_{i,0}+\slim_{j=1}^\mc{L} Q_{ij}\hat{u}_j(\ell x)\ell^{k_j+1}x^{k_j+1},\quad i=1,...,\mc{L},
\end{equation}
and working with the field equations in the form \eqref{EE12} -- \eqref{YM2}.

Substituting (\ref{l0var}, \ref{omega0}) into the field equations, again we find that $\check{m}(x)$ and $S(x)$ must be constant, which due to boundary conditions we are forced to set equal to 0 and 1 respectively. We also see that if $\ell=0$, all gauge functions $\omega_i(x)\equiv\omega_{i,0}$, and the solution reduces to the SadS case where $\omega_j\equiv\pm\lambda^{1/2}_j$, which are manifestly nodeless. However it is important to examine the behaviour of the equations for $\ell$ small but non-zero. 

When $\ell=0$, the YM equations \eqref{YM2} decouple to produce the following:
\begin{equation}
x(1+x^2)\frac{d^2\hat{u}_j}{dx^2}+2\left(k_j+(k_j+1)x^2\right)\frac{d\hat{u}_j}{dx}+xk_j(k_j+1)\hat{u}_j=0,
\end{equation}
where we have used results (\ref{BigF}, \ref{BigF2}, \ref{frewrite}).

Fortunately, though not necessarily unexpectedly, this is also very similar to the $\sun$ case \cite{baxter_existence_2008} (set $k_j\equiv k$ in the above) in that the term containing $\mc{F}$ vanishes in both cases when $\ell=0$. Therefore our more general case has a very similar unique solution in this limit:
\begin{equation}\label{solnl0sol}
\check{m}(x)\equiv 0,\qquad S(x)=1,\qquad\hat{u}_j(x)\propto\,_2 F_1 \left(\frac{k_j+1}{2},\frac{k_j}{2};\frac{2k_j+1}{2};-x^2\right)
\end{equation}
for $j=1,...,\mc{L}$, and where the integers $k_j$ for the group $G$ in question are given in Table \ref{Table1}. The constant of proportionality above is simply $\beta_j$ from Proposition \ref{prop:lex0}. It can be seen that this is regular at $x=0$, and due to the properties of hypergeometric functions, that it satisfies the required boundary conditions \eqref{BCinf}.

We proceed in a very similar fashion to the black hole case. Proposition \ref{prop:lexinf} adapts in a very obvious way, similar to the above (\ref{barm}, \ref{Xpm}). So we take the existing solution \eqref{solnl0sol} with arbitrary $\beta_j$, and consider varying $\{\beta_j\}$ and varying $\ell$ away from 0. Note again that for the embedded solution \eqref{solnl0}, all gauge fields will be nodeless. We then choose some $r_1 >> 0$ so that we can consider $r_1$ in the asymptotic regime. Propositions \ref{prop:lex0} guarantees that for fixed $\ell$ sufficiently small we can find solutions near the existing unique solution which will begin regularly near $r = 0$ and remain regular in the range $(0, r_1]$, and that those solutions will have nodeless gauge field functions due to analyticity. Finally, once we are in the asymptotic regime, we can again use Proposition \ref{prop:gloreg} and the logic in Section \ref{sec:asym} to ensure that solutions will remain regular as $r\rar\infty$, and that furthermore all these nearby $\omega_j$ will be nodeless. $\Box$

\section{Conclusions}\label{GGEConc}

The purpose of this research was to investigate the existence of global black hole and soliton solutions to spherically symmetric, four dimensional EYM theories with compact semisimple connected and simply connected gauge groups.

We began by stating the basic elements of the theory, describing the analogy to the asymptotically flat case considered in \cite{oliynyk_local_2002}. We derived the basic field equations for adS EYM theory, and then explained how to reduce the model down to the case for the regular action \cite{oliynyk_local_2002,brodbeck_generalized_1993}, in which the constant isotropy generator $W_0$ lies in an open fundamental Weyl chamber of the Cartan subalgebra $\mk{h}$. In this case it may be shown that the regular action reduces to the principal action described in \cite{dynkin_semisimple_1952}, which simplified the field equations greatly.

We went on to investigate the boundary conditions at $r=0$, $r=r_h$ and as $r\rar\infty$ (Section \ref{sec:BCs}). We found that the analysis at the event horizon and at the origin (Propositions \ref{prop:lex0} and \ref{prop:lexrh}) carried over similarly from the asymptotically flat case \cite{oliynyk_local_2002}, with some minor alterations. The biggest difference in the analyses was in the asymptotic behaviour of solutions (Proposition \ref{prop:lexinf}). There, we found that the gauge functions and their derivatives were entirely specified by the arbitrary values they approach at infinity -- this differs greatly from the $\Lambda = 0$ case, in which the gauge field was specified by higher order parameters in the power series, and these parameters were intercoupled in a complicated way. This difference is explained in Section \ref{sec:asym}, where it is noted that due to the parameter we use to render the equations autonomous, the solutions to this system (in terms of dynamical systems) need not reach their critical points, which was what forced the asymptotically flat system to be so tightly constrained as $r\rar\infty$.

Due to this difference, it became possible in Section \ref{sec:gloex} to prove the existence of global solutions to the field equations in some neighbourhood of embedded solutions, of which we found three separate cases (Section \ref{sec:embed}). We proved that as long as $\mu(r)>0$ throughout the solution range, then if we begin at the initial boundary ($r=r_h$ for black holes or $r=0$ for solitons) and integrate the field equations out arbitrarily far, the field variables will all remain regular (Proposition \ref{prop:gloreg}). We recall that we already established in Section \ref{sec:asym} that general solutions will remain regular in the asymptotic regime. Therefore, we were able to argue the existence of black hole and soliton solutions which begin regularly at their initial conditions and can be regularly integrated out arbitrarily far, where they will remain regular as $r\rar\infty$ (Theorem \ref{prop:gloex}). We finally considered the limit of $|\Lambda|\rar\infty$, which we explained was necessary in the $\sun$ case to guarantee nodeless and hence stable solutions, and proved that nodeless non-trivial solutions exist in this regime too, which are similarly globally regular and analytic in their boundary parameters (Theorem \ref{thr:l0}).

Our main results are the proof of global non-trivial solutions to the field equations \eqref{EE1} -- \eqref{YM}, both nearby trivial embedded solutions, and in the limit of $|\Lambda|$ large. It is remarkable to see how many of the general features of this model carry across to the specific case of $\sun$ \cite{baxter_existence_2008}. These include the forms of the field equations themselves, the embedded solutions we find, the qualitative behaviour of the solutions at the various boundaries, and the existence of solutions both near embedded solutions and in the limit $|\Lambda|\rar\infty$. This is very pleasing, since it may be noticed that the field equations \eqref{EE1} -- \eqref{YM} may easily be adapted to any gauge group without precise knowledge of the gauge potential itself, the construction of which for a given gauge group is a non-trivial task. This quite general system, even restricted to solely the regular case, could thus prove to be a powerful analytical model which may give insight into a range of different matter field theories.

There are many future directions that this work could take. Considering the work in \cite{kunzle_all_2002}, a logical next step might be to consider the `irregular' case, where $W_0$ lies on the boundary of a fundamental Weyl chamber, and the situation is more intricate. For instance, for $\Lambda=0$ it is known that this means the gauge functions $\omega_j$ will in general be complex. An analysis of that case, in combination with the results here presented, would cover an existence analysis for black holes and solitons in all possible static, spherically symmetric, purely magnetic EYM adS models with a compact semisimple gauge group.

Another obvious thing to do is to consider the question of the stability of the solutions that we have found. In \cite{brodbeck_instability_1996}, Brodbeck and Straumann give a proof of instability for a general compact gauge group in asymptotically flat space, for the case of the regular action; but here we find that we are able to establish solutions which fulfil the same conditions which guaranteed stability in the case of $\sun$. This would be very enlightening to investigate. 
In addition, there is the issue of extending this work to higher dimensions, though due to the fact that we would now be dealing with essentially $SU(3)$ principal bundle automorphisms for the isometry group of $S^3$, and the higher order Cherns-Simons terms in the action needed to obtain finite-mass solutions \cite{brihaye_particle-like_2003, brihaye_higher_2003}, this is likely to be highly  technical.

The main impact of this research is on some outstanding questions in gravitational physics. For instance, we consider Bizon's modified ``no-hair" theorem in light of this work, which states:

\begin{quote}
	``Within a given matter theory, a stable black hole is characterised by a finite number of global charges." \cite{bizon_gravitating_1994}
\end{quote}

Since this work concerns a general gauge group, it opens up the interesting possibility of verifying the no-hair theorem for a large class of gauge structure groups, given some further work. In addition, Hawking very recently raised the interesting possibility that hairy black holes may be used to resolve the `black hole information paradox' \cite{hawking_soft_2016}. The possibilities that this research opens up for our field are as yet unknown but potentially significant, and it would be of great interest to know if our recent work may be able shed any light on this long-standing problem.

Finally, there is the important question of whether this research will open up new insights into the adS/CFT correspondence. It is known that for black hole models there are observables in the dual CFT which are sensitive to the presence of hair (see \cite{hertog_black_2004} for a discussion of non-Abelian solutions in the context of adS/CFT), and correspondences to CMP problems have been found relating to both superconductors \cite{cai_introduction_2015, cai_black_1996} and superfluids \cite{sachdev_strange_2013}. Therefore, it is possible that within the class of models considered in this paper, there exist many more applications to QFT phenomena, and this could be a rich and worthwhile vein of study. 

%
\appendix

\section*{Acknowledgements}

The author would like to express great thanks to Dr T. Oliynyk (Monash University, Melbourne, Australia) for a very useful email exchange.

\vspace{1cm}
\hrule

\section*{References}
%
%
%
\bibliographystyle{unsrt}


\end{document}